\documentclass[aip,nofootinbib,12pt]{revtex4}
\usepackage{graphics}
\usepackage{bm}
\newcommand{\Jos}{Josephson }
\newcommand{\jun}{junction }

\usepackage{amsmath}    
\usepackage{amssymb}    
\usepackage{graphicx}   
\usepackage{epsf}
\usepackage{epsfig}
\usepackage[hang]{subfigure}
\begin{document}

\title {Static Properties of Small Josephson Tunnel Junctions in a Transverse Magnetic Field}
\thanks{Submitted to Journal of Applied Physics}

\author{R.\ Monaco}
\affiliation{Istituto di Cibernetica del C.N.R., 80078, Pozzuoli, Italy and Unita' INFM-Dipartimento di Fisica, Universita' di Salerno, 84081 Baronissi, Italy.}\email
{roberto@sa.infn.it}
\author{M.\ Aaroe}
\affiliation{Department of Physics, B309, Technical University of
Denmark, DK-2800 Lyngby, Denmark.} \email{aaroe@fysik.dtu.dk}
\author{V.\ P.\ Koshelets}
\affiliation{Institute of Radio Engineering and Electronics, Russian
Academy of Science, Mokhovaya 11, Bldg 7, 125009, Moscow,
Russia.}\email{valery@hitech.cplire.ru}
\author{J.\ Mygind}
\affiliation{Department of Physics, B309,  Technical University of
Denmark, DK-2800 Lyngby, Denmark.} \email{myg@fysik.dtu.dk}

\date{\today}

\begin{abstract}
The magnetic field distribution in the barrier of small planar
Josephson tunnel junctions is numerically simulated in the case when
an external magnetic field is applied perpendicular to the barrier
plane. The simulations allow for heuristic analytical solutions for
the Josephson static phase profile from which the dependence of the
maximum Josephson current on the applied field amplitude is derived.
The most common geometrical configurations are considered and, when
possible, the theoretical findings are compared with the
experimental data.
\end{abstract}


\maketitle
\newpage
\section{Introduction}

The static (and dynamic) properties of a planar Josephson Tunnel
Junction (JTJ) are well understood when an external magnetic field
is uniformly applied in the junction plane \cite{barone}. On the
contrary, very little is known when a uniform magnetic field is
applied perpendicularly to the barrier plane. The main reason why,
since the discovery of the Josephson effect in 1962, only few papers
have dealt with a transverse magnetic
field\cite{rc}\cite{hf}\cite{miller}, is due to the fact that
demagnetization effects imposed by the electrodes geometry are
awkward to take into account. In a recent paper \cite{mon} we
provided an experimental proof that a transverse magnetic field can
be much more capable than an in-plane one to modulate the critical
current $I_c$ of a planar JTJ with proper barrier and electrodes
geometry requirements. It is also possible to design the JTJ
electrode geometry in such a way that it is totally insensitive to a
transverse field. The possibility to have on the same chip JTJs
having different sensitivities to an externally applied field can be
very attractive in practical applications.

\noindent In this paper we push our analysis further by resorting to
numerical magnetostatic simulations to find the field distribution
${\bf H}$ in the barrier plane of those JTJs having the most common
rectangular and annular geometries. Once ${\bf H}$ is found
empirically, the Josephson phase $\phi$, which is the difference
between the complex wavefunction phases in the electrodes, can be
obtained from the Josephson equation\cite{joseph}:

\begin{equation}
\label{gra}
{\bf \nabla} \phi = \frac{2\pi d_e \mu_0}{\Phi_0}{\bf H}\times {\bf n} ,
\end{equation}

\noindent where ${\bf n}$ is a unit vector normal to the insulating
barrier separating the two superconducting electrodes, $\mu_0$ is
the vacuum permeability and $\Phi_0=h/2e$ is the magnetic flux
quantum. If the two superconducting films have thicknesses $d_{1,2}$
and London penetration depths $\lambda_{L1,2}$ and $t_j$ is the
barrier thickness, then the effective magnetic penetration $d_e$ is
given by\cite{wei}:

\begin{equation}
d_e=t_j + \lambda_{L1} \tanh {d_1 \over 2 \lambda_{L1}} + \lambda_{L2} \tanh {d_2 \over 2 \lambda_{L2}},
\label{d_e}
\nonumber
\end{equation}

\noindent which, in the case of thick superconducting films ($d_i
>> \lambda_{Li}$), reduces to $d_e \approx \lambda_{L1} +
\lambda_{L2}$ (being always $d_i >> t_j$).

\noindent In Cartesian coordinates, assuming that the tunnel barrier
lies in the $x-y$ plan, Eq.\ref{gra} reduces to:

\begin{equation}
\label{pds} \frac{\partial \phi(x,y)}{\partial x}\propto
-H_y,\,\,\,\,
 \frac{\partial \phi(x,y)}{\partial y}\propto H_x.
\end{equation}

\noindent For a planar JTJ with a uniform Josephson current density
$J_{c}$ whose dimensions are smaller than the Josephson penetration
depth $\lambda _{J}=\sqrt{\hbar /2e\mu _{0}d_{e} J_{c}}$, the
self-induced field associated with the bias current can be neglected
and the Josephson phase must satisfy the two-dimensional Laplacian
equation \cite{joseph}:

\begin{equation}
\label{laplace} \frac{\partial^2 \phi}{\partial x^2} +
 \frac{\partial^2 \phi}{\partial y^2}=0,
\end{equation}

\noindent with proper boundary conditions related to the value of
the magnetic field components $H_x$ and $H_y$ on the junction
perimeter. It was first pointed out in 1975\cite{hf} that in a
transverse applied field $\bf{H}= \rm{H_\bot} \bf{\hat{z}}$, the
in-plane components $H_x$ and $H_y$ are ascribed to surface
demagnetizing currents $\bf{j_s} = \bf{\hat{z}} \times \bf{H}$
feeding the interior of the junction. Since these currents mainly
flow on the film edges, the largest sensitivity to a transverse
field occurs when the junction is formed at the film edges. On the
contrary, if the barrier is placed well inside the superconducting
films, the effect of a transverse field vanishes. Our task consists
of numerically evaluating the field line distribution in the barrier
plane, from which we determine an empirical analytical expression
$\phi(x,y)$ for the phase profile which satisfies Eq.\ref{laplace}.
Such a phase profile will allow the computation of the transverse
magnetic diffraction pattern $I_c (H_\bot)$ for small JTJs having
different geometries and to compare it with experimental data, if
available. This is achieved by recalling that the maximum Josephson
current is:

$$I_c= I_0 \sqrt{ \langle \sin \phi \rangle^2 + \langle \cos \phi \rangle^2},$$

\noindent in which the brackets $\langle\rangle$ denote spatial
averages over the junction area. Throughout the paper we assume that
the applied transverse field is everywhere much smaller than the
critical field which would force the films into the intermediate or
normal state, i.e., that the superconductors are always in the
flux-free Meissner regime.

\section{Magnetostatic simulations}

In general, magnetostatic problems are based on the magnetic vector
potential. However, where no electrical currents are present, the
problem can be conveniently solved using the scalar magnetic
potential. In fact, in a current free region $\nabla \times
\bf{H}=0$ allows the introduction of a scalar potential $V_m$ such
that $\bf{H}= - \bf{\nabla} \rm{V_m} $. Using the constitutive
relation $\bf{B}= \mu_0 \mu_r \bf{H}$, we can rewrite Maxwell's
equation $\bf{\nabla} \cdot \bf{B}=0 $ in terms of $V_m$:

\begin{equation}
\label{Vm} -\bf{\nabla}\cdot (\mu_0 \mu_r \bf{\nabla} V_m ) =0,
\end{equation}

\noindent in which the magnetic relative permittivity $\mu_r$ is
spatially dependent. We assumed that the superconducting electrodes
are thicker than their London penetration depths ($d_i>>
\lambda_{Li}$), so that the London equation reduces to $\bf{B}=0$
everywhere inside the superconductors, i.e., $\mu_r= 0$ (perfect
diamagnetism) and the normal component of the magnetic flux density
vanishes at the boundary ($\bf{n} \cdot \bf{B} =0$). In the opposite
limit, the films would become transparent to the transverse field
and, in turn, the junction would lose its sensitivity to the
transverse field. A uniform applied magnetic field $\bf{H}=
\rm{H_\bot} \bf{\hat{z}}$ is taken into account by imposing that
sufficiently far away from the junction is $V_m= - H_\bot z$. All
the simulations presented in this paper were carried out setting
$H_\bot = 1A/m$.

\noindent As a consequence of the definitions of Eq.\ref{pds}, it is
straightforward to show that Eq.\ref{laplace} requires that
$\partial^2 V_m / \partial x
\partial y = \partial^2 V_m / \partial y \partial x$.
Further, more importantly, we have:

\begin{equation}
\label{Vmm} \phi(x,y) \propto \int dx \frac{\partial V_m}{\partial
y} = - \int dy \frac{\partial V_m}{\partial x}.
\end{equation}

\noindent The numerical solution of Eq.\ref{Vm} was implemented in
the COMSOL Multiphysics 3D Electromagnetics module for JTJs having
different rectangular and annular geometries. Models with large
geometric scale variations are always problematic to mesh, in
particular if they contain thin layers with large aspect ratio.
Therefore, one caveat of our modeling is that, in order to keep the
number of mesh elements within the PC memory handling capability,
the separation between the superconducting films, i.e. the tunnel
barrier thickness $t_j$, could not be set to realistic values for a
Josephson tunnel barrier $O(1nm)$. Our numerical modeling was tested
against the magnetic field distribution around a superconducting
disk (with radius $R$ larger than its thickness $d$) in the plane
$z=0$, centered on the $z$ axis and immersed in a field $\bf{H}=
\rm{H_\bot} \bf{\hat{z}}$. More precisely, the radial dependence of
the tangential field $H_t$ on the disk surface followed to a high
accuracy the well known expression\cite{landau} $H_t(r) \propto
H_{\bot} r/ \sqrt{R^2-r^2}$ everywhere except at the disk border,
where the inverse square root singularity was replaced by a finite
value $\hat{H}$ proportional to the square root of the disk aspect
ratio $\hat{H}=H_\bot \sqrt{R/d}$\cite{benk}. This example is
indicative of the fact that, in general, the magnetostatic response
of any superconducting film structure is markedly dependent on the
film aspect ratio.

\begin{figure}
\begin{center}

\epsfysize=6.0cm \epsfbox{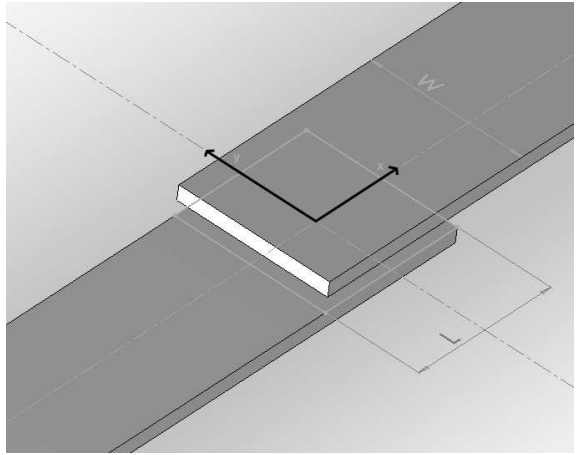}
\end{center}
\caption{Sketch of a overlap geometry junction . The center of the
junction coincides with origin of our coordinate system.}
\label{overgeom}
\end{figure}

\begin{figure}
\centering

\subfigure[]{\includegraphics[width=9cm]{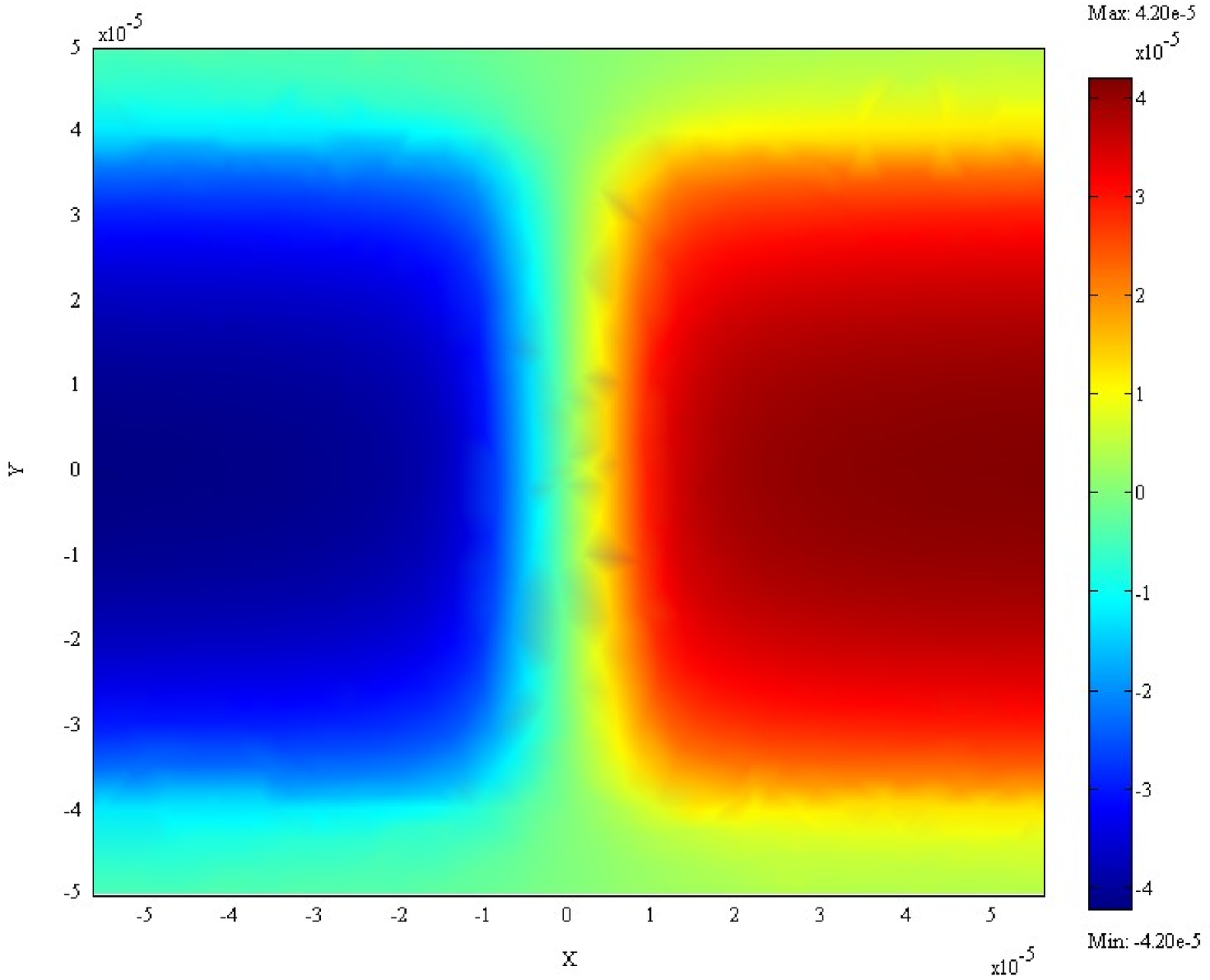}}
\subfigure[]{\includegraphics[width=9cm]{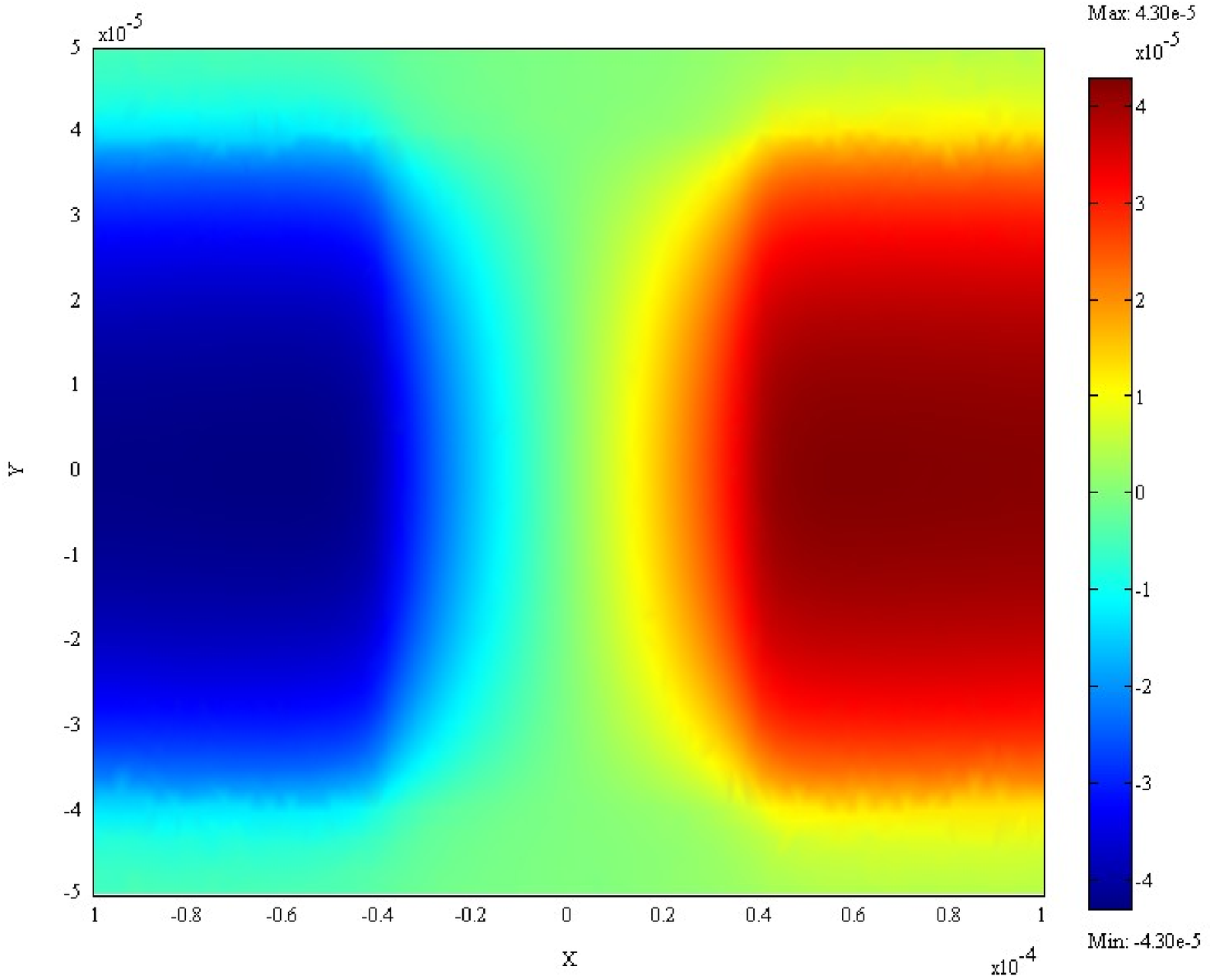}}
\subfigure[]{\includegraphics[width=9cm]{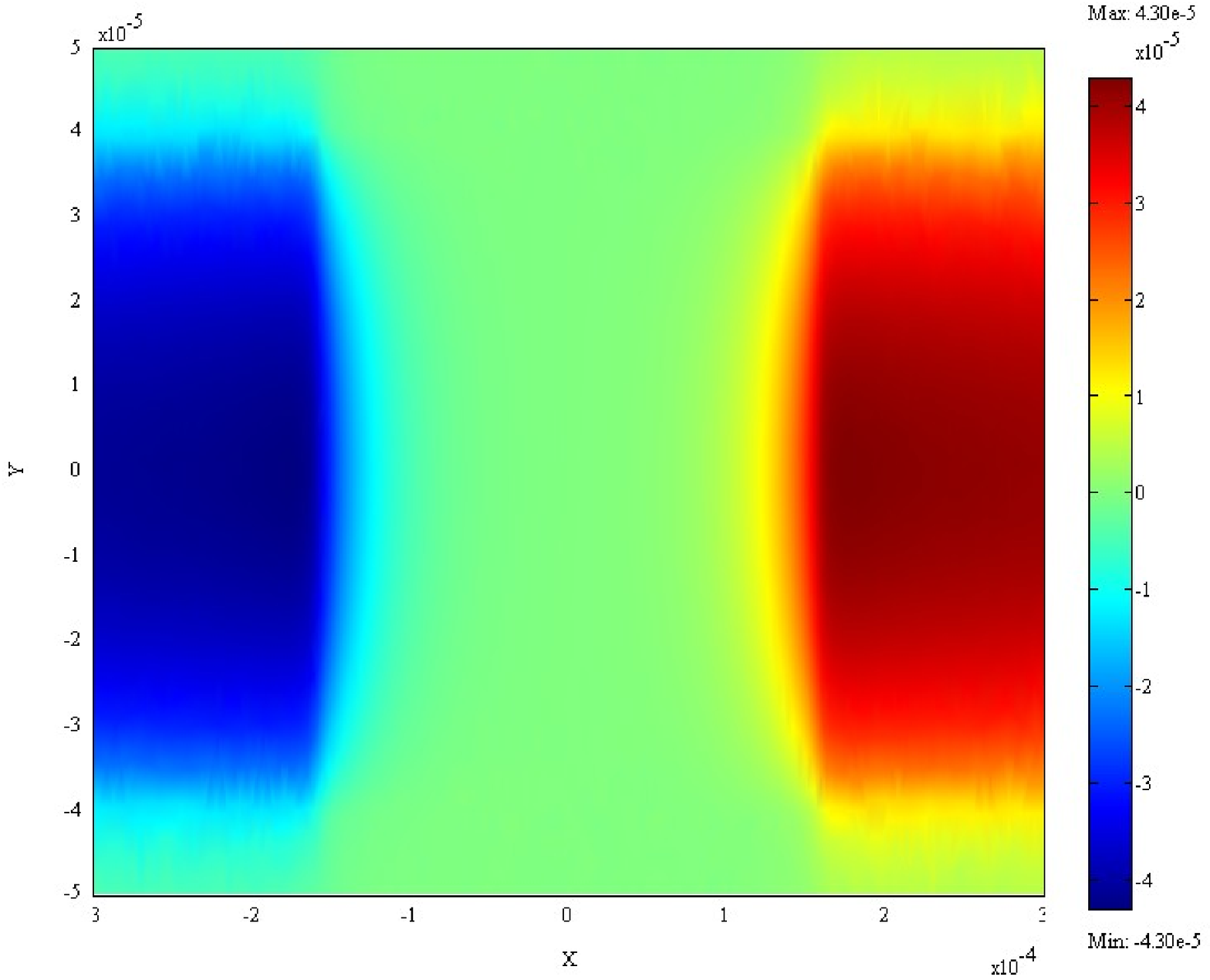}}
\caption{(Color online) Numerically obtained magnetic scalar
potential $V_m$ (in $A$) for three overlap planar Josephson tunnel
junctions having the same width $2W=80\mu m$, but different lengths:
a) $2L=20\mu m$ ($\beta=0.25$), b) $2L=80\mu m$ ($\beta=1$), and
 c) $2L=320\mu m$ ($\beta=4$). The external applied field is $H_\bot=1A/m$.}
\label{Vm over}
\end{figure}

\section{Rectangular junctions}

\subsection{Overlap type junctions}

\noindent We begin our analysis with a JTJ obtained by the
superposition of the extremities of two long and narrow parallel
superconducting electrodes with equal widths. This so-called
\textit{overlap} geometry is depicted in Fig.\ref{overgeom} for a
square junction, i.e. $W=L$. The tunnel barrier lies in the $z=0$
plane and its center coincides with the axis origin. Further, it has
a length $2L$ along the x-direction and a width $2W$ along the
y-direction. In the simulations the electrodes have a thickness
$d=10\mu m$ and are $t_j=1\mu m$ apart. The film width $2W$ and the
overlapping length $2L$ where varied in order to treat barriers with
different aspect ratios $\beta=L/W$. Figs.\ref{Vm over}a-c show the
numerically obtained $V_m$ solutions in the barrier area of three
overlap junctions having the same width $2W=80\mu m$, but different
lengths $2L=20$, $80$, and $320\mu m$. By analyzing the properties
of such plots we aim to infer an empirical, physically acceptable
analytical form for $V_m(x,y,z=0)$. We observe that, for any value
of $\beta$, the scalar potential in the barrier is symmetric with
respect to the x-axis and antisymmetric with respect to the y-axis.
In other words, the expression $V_m(x,y)$ we are looking for has to
be an odd function of $x$ and a even function of $y$. Furthermore,
we note that the potential decays from the junction corners over a
distance $W$, being mostly null when $L>W$ (or $\beta>1$). We have
checked that the following ansatz:

\begin{equation}
\label{Vmover} V_m(\hat{x}, \hat{y}) = W \hat{H} \cos q
\hat{y}\frac{\sinh q \beta \hat{x}}{\sinh\beta},
\end{equation}

\noindent in which $q$ is a fitting parameter near unity, allowed us
to reproduces the plots of Figs.\ref{Vm over}a-c at a better than
qualitative level. In fact, for $q=1$ the relative difference
between the simulation output and the proposed expression being
everywhere less than $\pm 15 \%$ and the $q$ value that minimized
the error was $q\simeq0.9$. We have introduced the normalized units
$\hat{x}=x/L$ and $\hat{y}=y/W$ (note that $\beta \hat{x}= x/W$). In
the last equation, again $\hat{H}\propto H_{\bot}\sqrt{W/d}$, with a
proportionality constant of order of unity which slightly increases
when the barrier thickness $t_j$ decreases. Unfortunately, recalling
the comments of the previous section, we cannot be more precise on
this point.

\begin{figure}
\centering

\subfigure[]{\includegraphics[width=8cm]{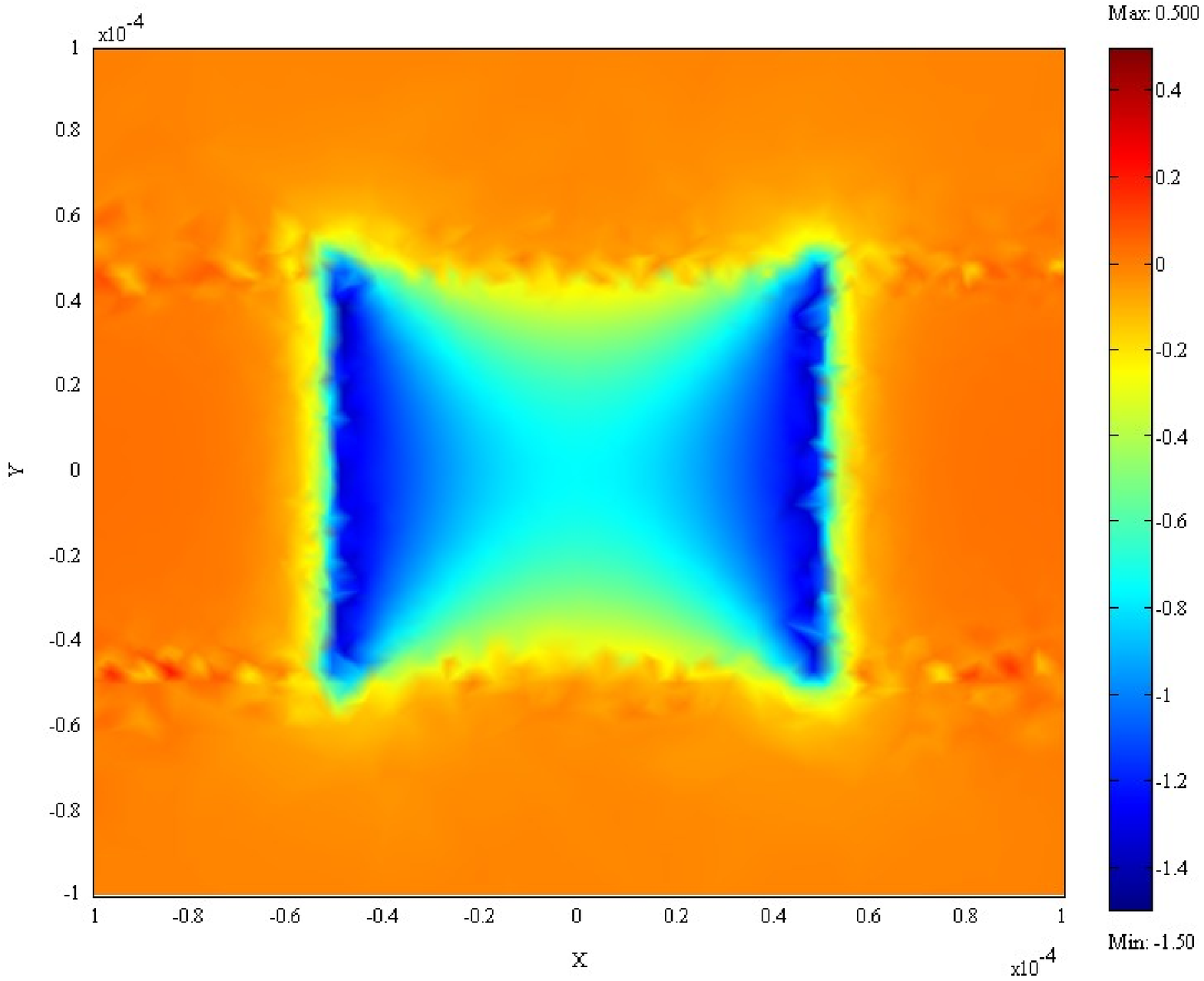}}
\subfigure[]{\includegraphics[width=8cm]{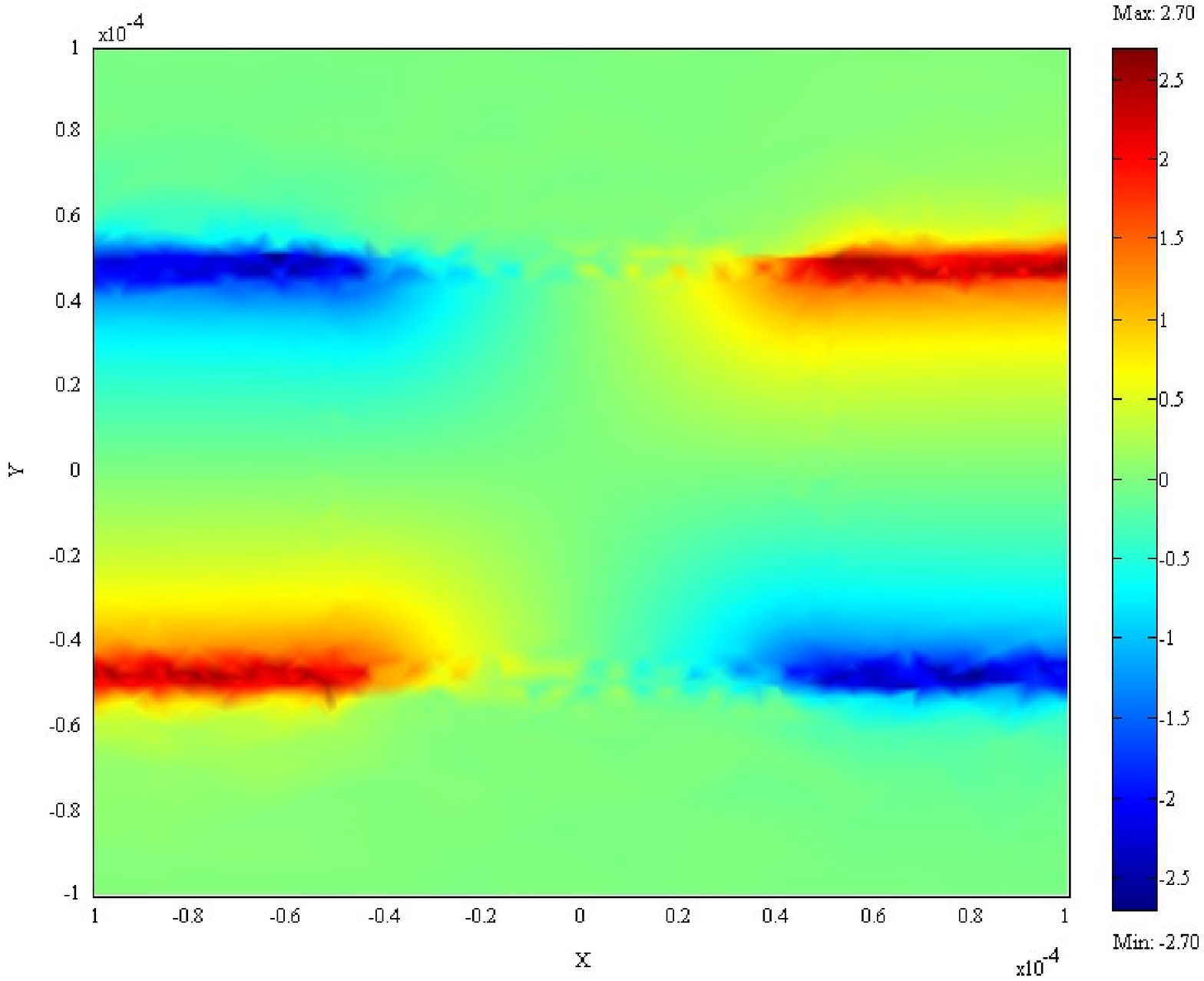}}

\caption{(Color online) Numerically obtained in-plane magnetic field
components in $A/m$ for a square overlap junction having
$2W=2L=100\mu m$ in a transverse externally applied field
$H_\bot=1A/m$. Color plots for a) $H_x(x,y,z=0)$ and b)
$H_y(x,y,z=0)$. } \label{over1}
\end{figure}

\noindent Now we focus our attention on the components of the
magnetic field in the barrier plane $H_x (x,y)=-\partial V_m
/\partial x|_{z=0}$ and $H_y (x,y)=-\partial V_m /\partial y|_{z=0}$
($H_z$ being identically null all over the barrier area). They are
shown in Fig.\ref{over1}a and b, respectively, for the particular
case $W=L=100\mu m$. We like to specify, at this point, that the
same plots obtained from numerical simulations based on the vector,
rather than scalar, potential differed by no more than $\pm 10 \%$,
the discrepancy being larger at the barrier edges. From
Eq.\ref{Vmover} with $q=1$, the following analytical expressions are
derived:

\begin{eqnarray}
\label{H_x} H_x(\hat{x},\hat{y})= -\hat{H}\cos\hat{y}\frac{\cosh\beta \hat{x}}{\sinh\beta},\\
\label{H_y} H_y(\hat{x},\hat{y})=\hat{H} \sin\hat{y}\frac{\sinh\beta
\hat{x}}{\sinh\beta}.
\end{eqnarray}

\begin{figure}[h]
\centering
\subfigure[]{\includegraphics[width=5cm]{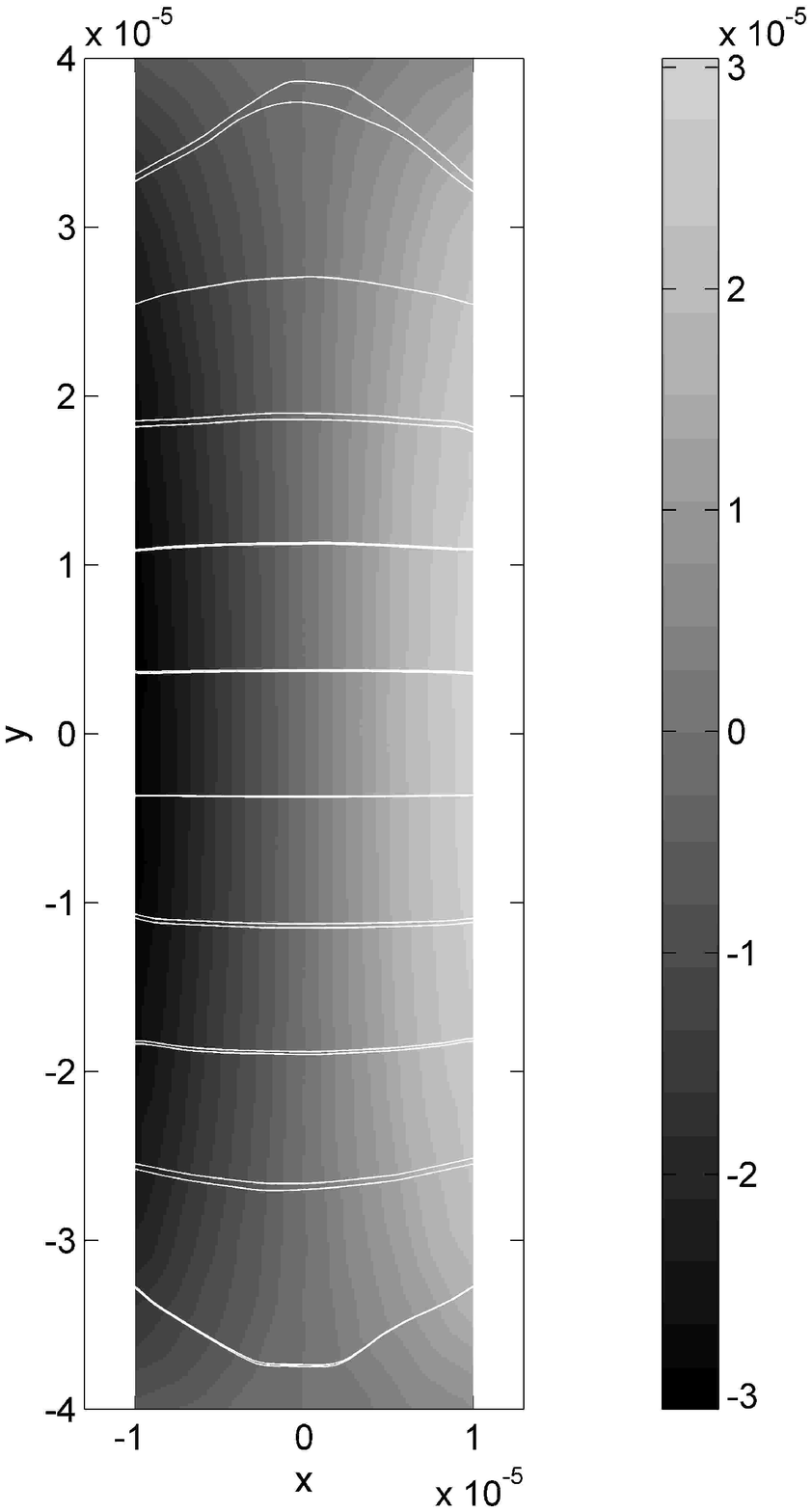}}
\subfigure[]{\includegraphics[width=7cm]{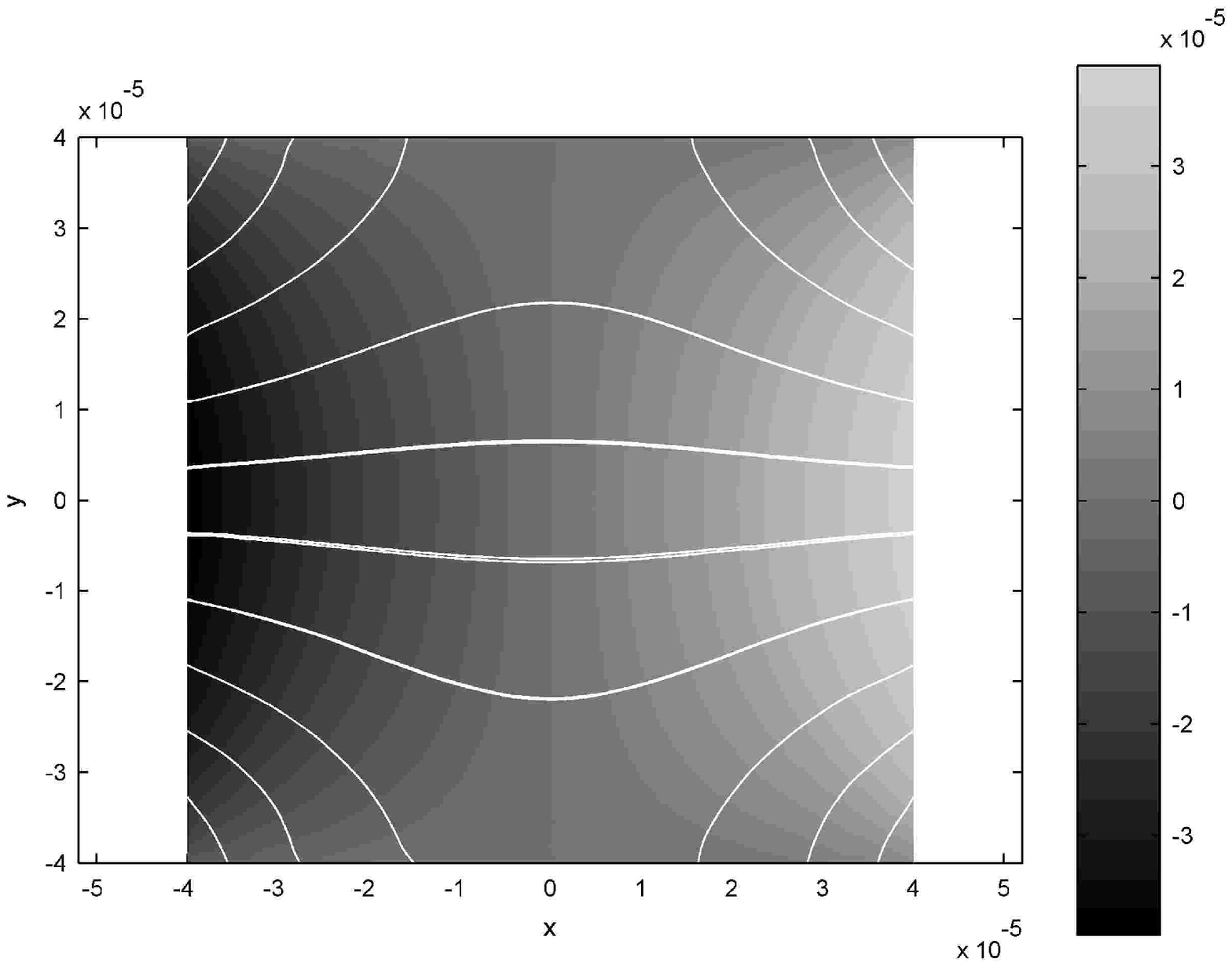}}
\subfigure[]{\includegraphics[width=7cm]{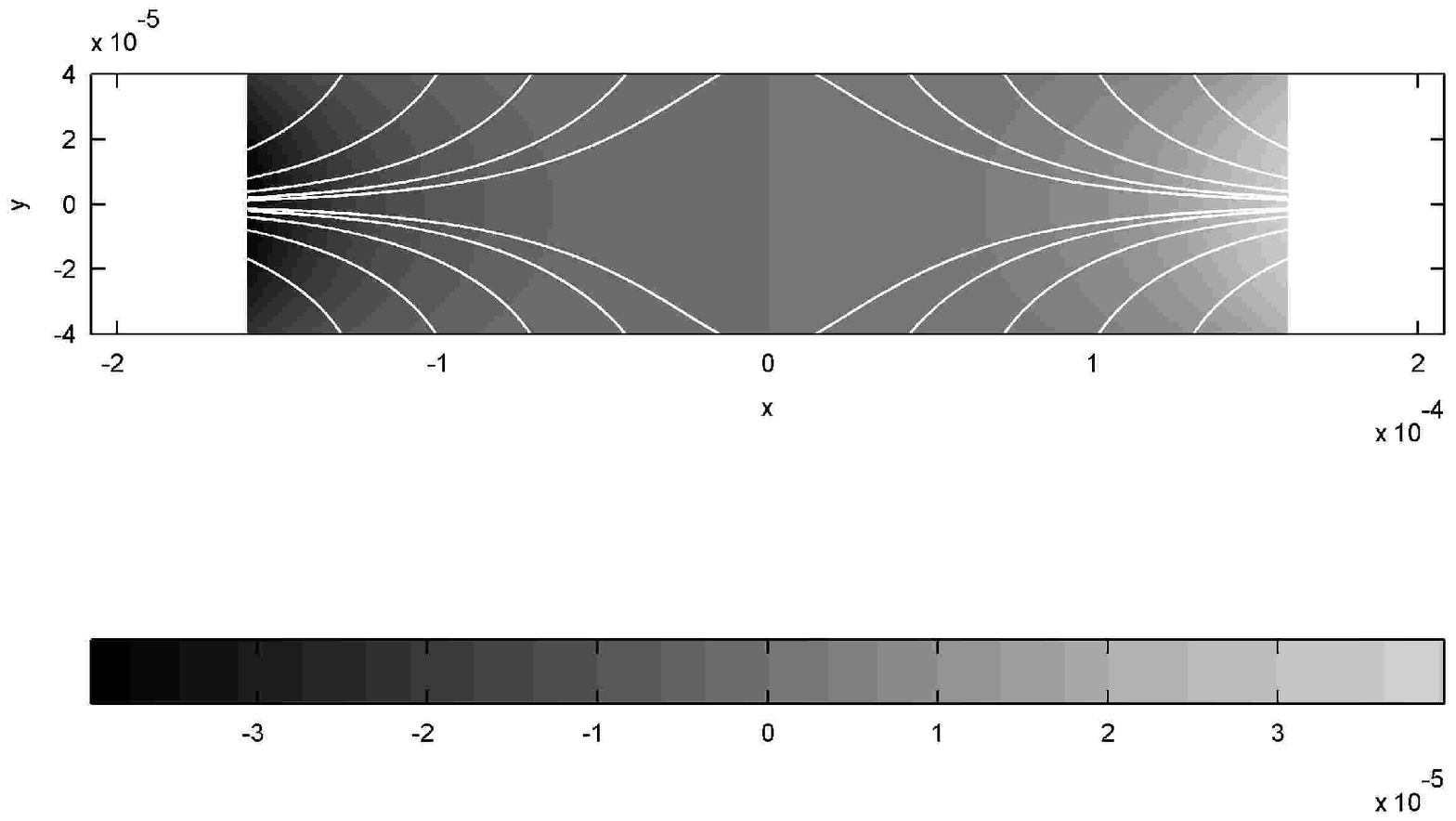}}
\caption{Magnetic field lines inside three overlap planar Josephson
tunnel junctions having the same width $2W=80\mu m$, but different
lengths: a) $2L=20\mu m$ ($\beta=0.25$), b) $2L=80\mu m$ ($\beta=1$)
and c) $2L=320\mu m$ ($\beta=4$).} \label{Fig4}
\end{figure}

\noindent The physical meaning of the last expressions is that for
$\beta=L/W>1$, the magnetic field lines are confined to the corners
of the junctions at a distance $W$ and most of the field lines
entering the junction at $x=\pm L$ are bent by $90^o$ and leave at
$y=\pm W$.  In the opposite limit, $\cosh \beta x \approx1$, so the
x-dependence of $H_x$ disappears, meaning that all the field lines
entering the barrier at, say, $x=- L$ exit at $x=L$ (or viceversa).
Further, we notice that while the $x$-component is negative all over
the barrier area, the $y$-component symmetrically spans from
negative to positive values. Due to the linearity of Eq.\ref{Vm} e
the system symmetry with respect to the $z=0$ plane, if the
direction of the transverse field is reverted, then $H_x$ and $H_y$
simply invert their sign. The magnetic field line distributions in
the junction barrier corresponding to the scalar potentials of
Figs.\ref{Vm over}a-c are shown in Figs.\ref{Fig4}a-c. Similar plots
based on the previous analytical expressions would be practically
undistinguishable at the picture resolution level, therefore, they
will not be shown. From the magnetic field distributions we expect
that, for a given junction area $L x W$, the critical current $I_c$
of a planar JTJ with pure overlap geometry ($L<W$) modulates much
faster than that of a sample with pure in-line geometry ($L>W$). At
a first sight, it might seem that the effect of a transverse field
is qualitatively similar to that of an in-plane field applied along
the film direction, i.e. along the $x$-axis, in our case. However,
this is not true at a quantitative level because, in general,
$H_x(x, \pm W)$ is not constant in a transverse field.

\noindent Inserting Eq.\ref{Vmover} in any of the Eq.\ref{Vmm}, we
derive an approximate analytical expression for the Josephson phase
profile:

\begin{equation}
\label{phyover} \phi(\hat{x}, \hat{y}) = h
\sin\hat{y}\frac{\cosh\beta \hat{x}}{\sinh\beta},
\end{equation}

\noindent where $h=2\pi d_e W \mu_0 \hat{H} /\Phi_0$ is a
dimensionless parameter proportional to the applied transverse field
amplitude $H_{\bot}$ through $\hat{H}$. It is easy to verify that
the last expression, in which we have omitted an integration
constant $\phi_0$, satisfies Eq.\ref{laplace}.

\noindent With $\phi$ an odd function of $\hat{y}$, then $\langle
\sin \phi \rangle=0$; therefore, the magnetic pattern $I_c(h)$
reduces to:

\begin{equation}
\label{overpatt} I_c(h)=I_c(0)\int_{0}^{1} d\hat{x} \int_{0}^{1}
d\hat{y} \cos \phi(x,y).
\end{equation}

\begin{figure}[ht]
\centering

\subfigure[]{\includegraphics[width=5.5cm]{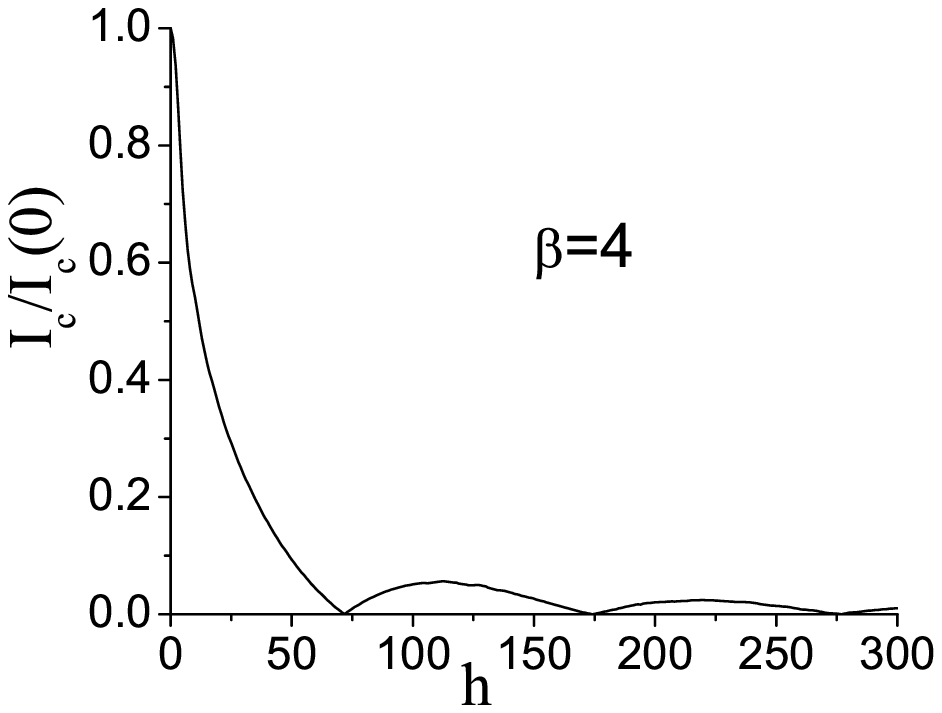}}%
\subfigure[]{\includegraphics[width=5.5cm]{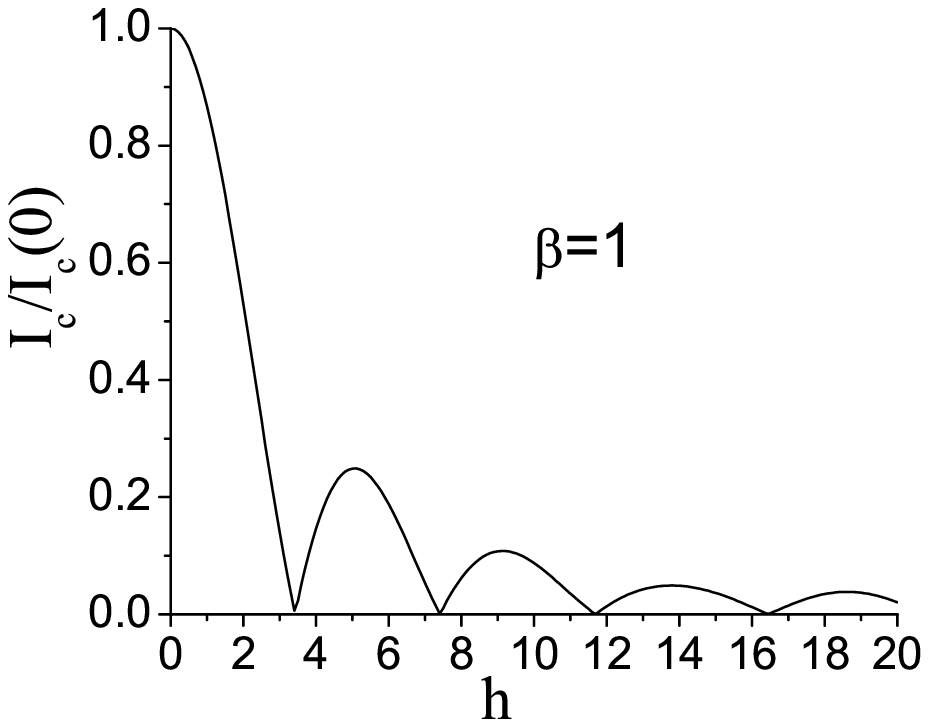}}%
\subfigure[]{\includegraphics[width=5.5cm]{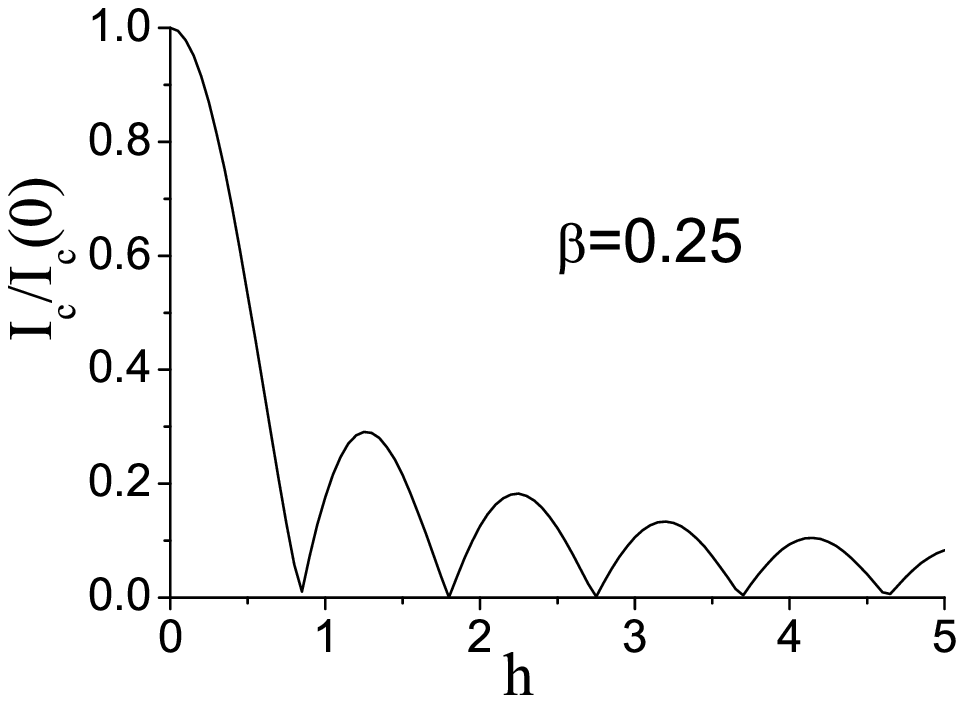}}%

\caption{Computed transverse magnetic patterns $I_c(h)$ for an
overlap junction with different $L/W$ ratios: a) inline junction
$L=4W$, b) square overlap $L=W$ and c) pure overlap junction
$W=4L$.} \label{pattover}
\end{figure}

\noindent Fig.\ref{pattover}a-c show the computed transverse
magnetic patterns for the three values of the barrier aspect ratio
$\beta$ used before ($4$, $1$ and $0.25$). As expected the response
to a transverse field is very weak for an in-line JTJ, the first
minimum occurring at $h\simeq 74$ for $\beta =4$. The secondary
pattern maxima become more pronounced for a pure overlap geometry.
However, in the limit $L<<W$, all the above equations lose their
validity when the overlapping length becomes comparable with the
film thickness.

\begin{figure}[h]
\centering

\subfigure[]{\includegraphics[width=8cm]{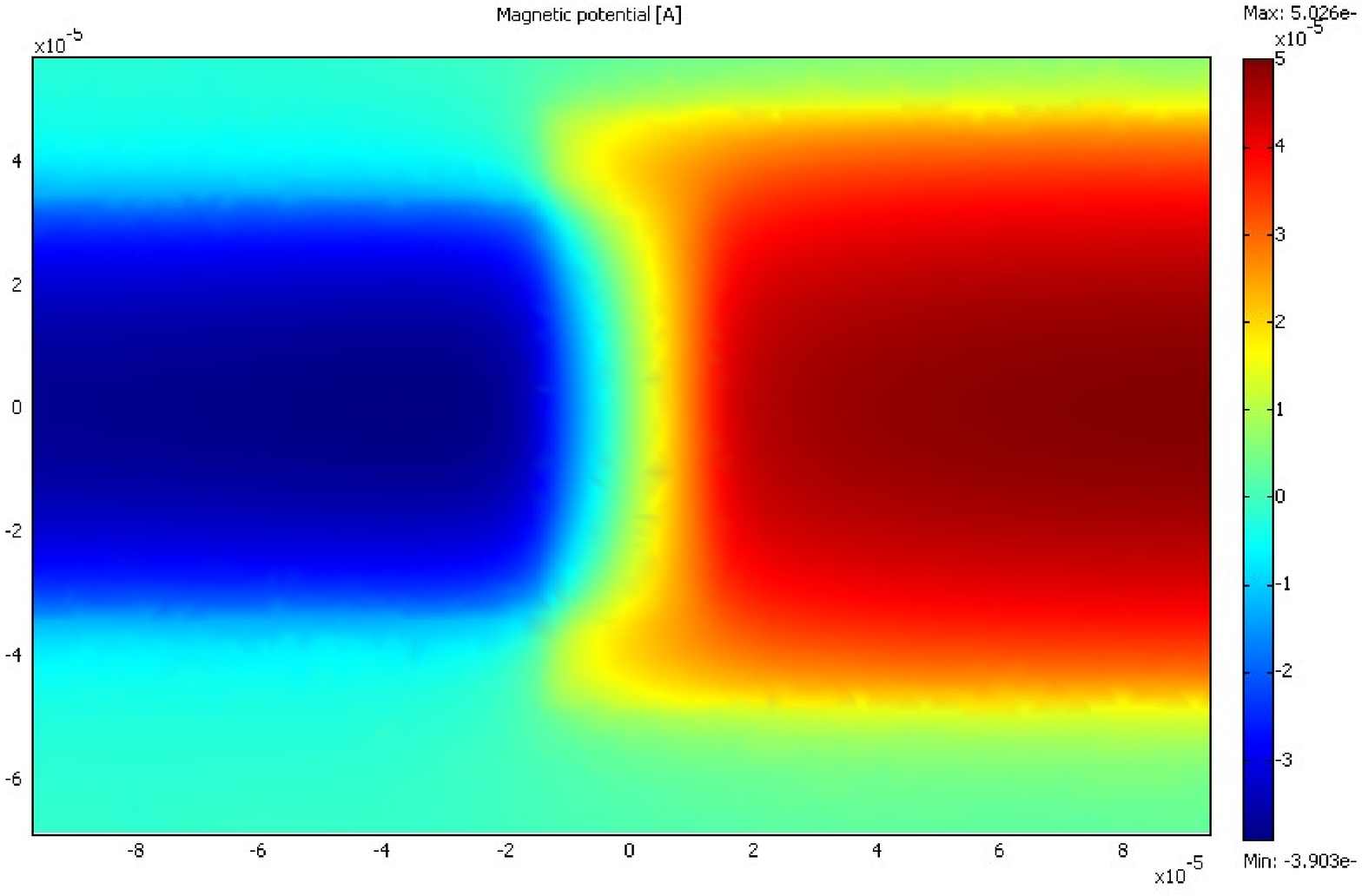}}
\subfigure[]{\includegraphics[width=8cm]{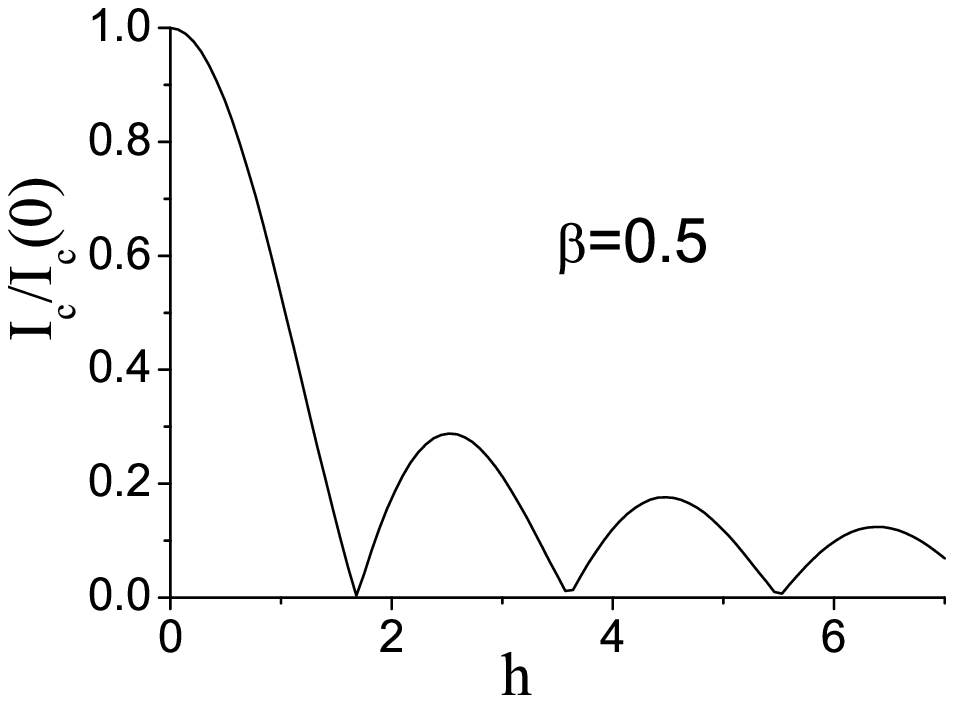}}

\caption{(Color online) a) Numerically obtained scalar potential for
an overlap junction obtained by the superposition of two films of
unequal widths $2W=0.74 mm$ and $2W'=1.00mm$. The overlapping
distance is $2L=0.35 mm$, as for the sample quoted in Ref.\cite{rc}.
b)Computed magnetic pattern $I_c(h)$ for an overlap junction with
aspect ratio $L/W=0.5$.} \label{rc}
\end{figure}

\noindent It is important to stress here that we are dealing with
electrically small JTJs, therefore the different shapes of the
transverse magnetic pattern is a direct consequence of the different
distribution of the surface screening currents (and not of the
applied bias current). Unfortunately, there are no data available in
the literature to check the validity of our theoretical magnetic
diffraction patterns for a small overlap JTJ formed by films having
the same widths. In fact, the experiments reported by Rosestein and
Chen in 1975\cite{rc} refer to an overlap JTJ formed by two thick
$Pb$ electrodes of unequal widths ($2W=0.74 mm$ and $2W'=1.00mm$)
and a common overlay region of $2L=0.35mm$. It is quite evident
that, for such geometrical film configuration, the symmetry with
respect to the $y$-axis is broken and Eq.\ref{phyover} is unable to
correctly describe the magnetic field (and screening currents)
distribution. Fig.\ref{rc}a and b show, respectively, the result of
numerical simulations carried out for the specific electrode
configuration of Ref.\cite{rc} and the corresponding $I_c(h)$.
According to Ref.\cite{hf}, we believe that difference between the
experimental data of Ref.\cite{rc} and the numerical prediction of
Ref.\cite{hf} valid only for the specific case $\beta=0.5$, arises
from the unequal widths of the films in the experiment. Indeed, the
magnetic diffraction pattern reported in Ref.\cite{hf} is of a piece
with the curve in Fig.\ref{rc}b.

\begin{figure}[h]
\centering
\subfigure[]{\includegraphics[width=8cm]{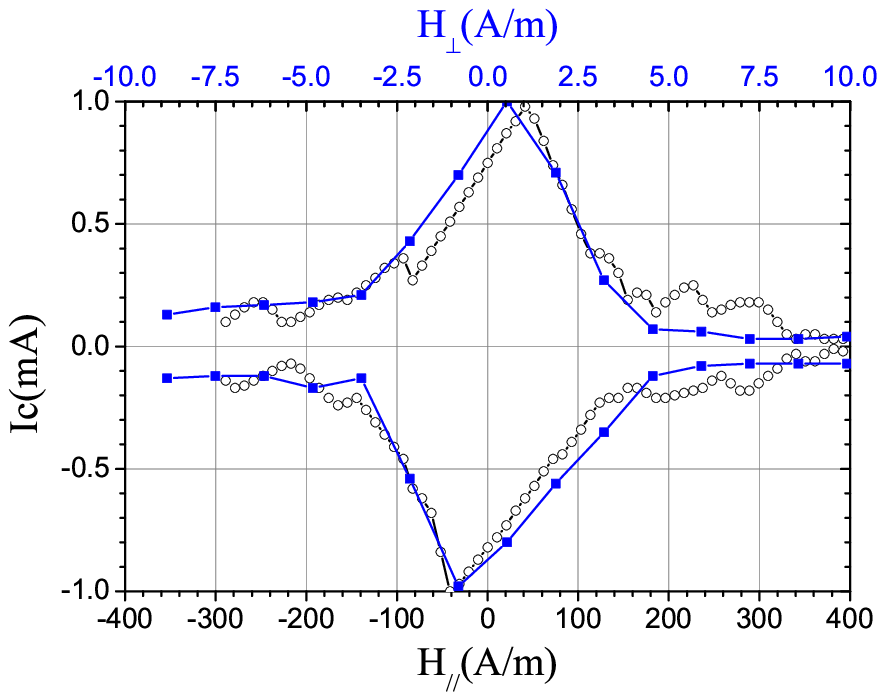}}
\subfigure[]{\includegraphics[width=8cm]{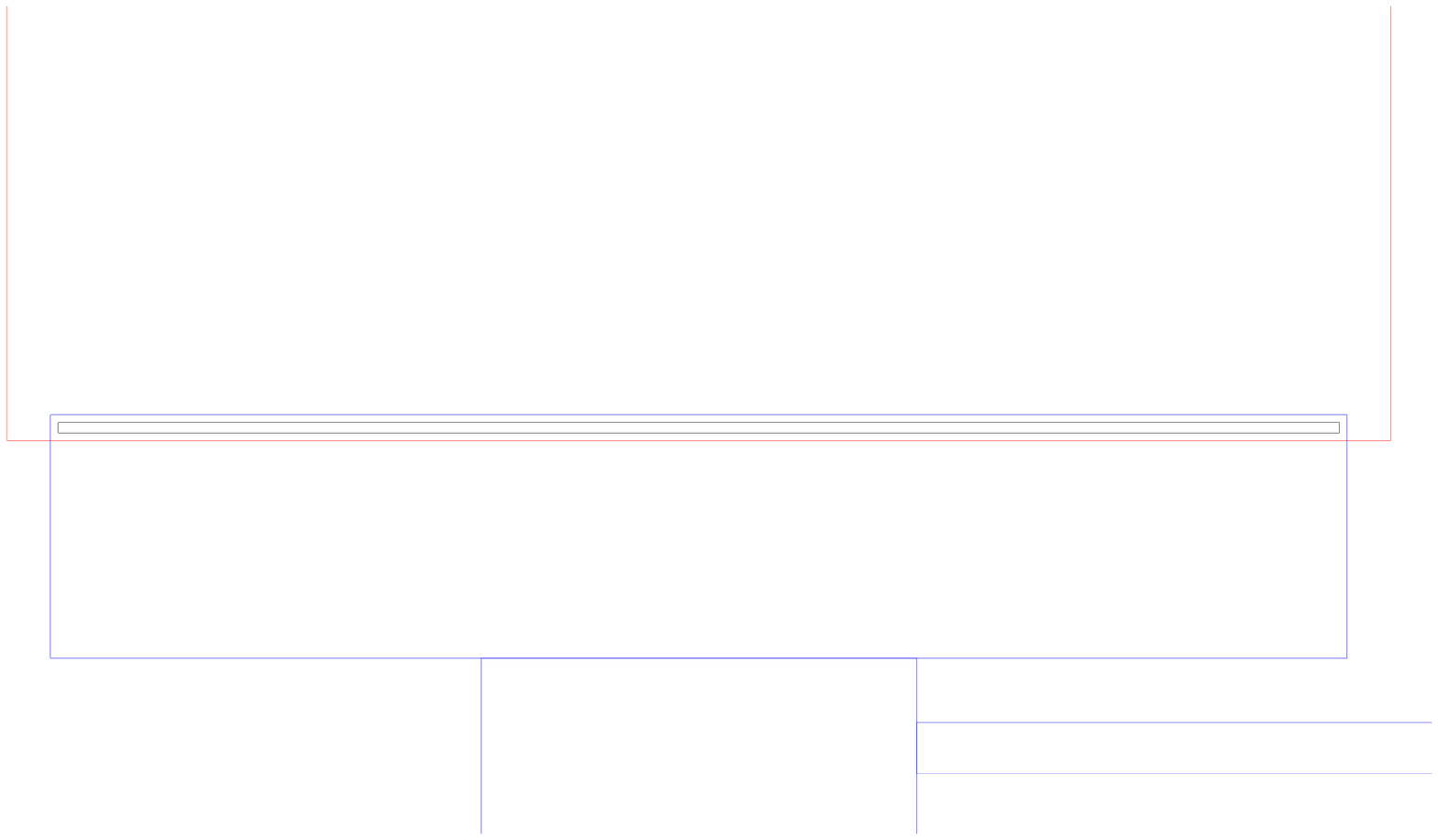}} \caption{a)
(Color online) Comparison between the diffraction patterns measured
in a parallel (black bottom axis) and transverse (blue top axis)
field of a $Nb/Al_{ox}/Nb$ overlap-type junction with $\lambda_J
\sim 50 \mu m$ whose length is $500\mu m$, while the width is equal
to $4\mu m$. b)Geometry details: the base (red) and top (blue)
electrode widths are $540$ and $506\mu m$, respectively. The barrier
area is delimited by the black rectangle.} \label{overexp}
\end{figure}

\noindent We conclude this section considering that, for
unidimensional overlap junctions for which $W>\lambda_j>L$, being
$\frac{\partial^2 \phi}{\partial x^2} << \frac{\partial^2
\phi}{\partial y^2}$, then the Josephson phase has to obey to the
equation first introduced by Owen and Sacalapino\cite{owen}:

\begin{equation}
\label{owen}  \frac{d^2 \phi}{dy^2}=\frac{1}{\lambda_j^2} sin \phi
\end{equation}

\noindent when an in-plane external field is applied along the
x-direction. In fact,  Fig.\ref{overexp}a shows the comparison
between the diffraction patterns measured in a parallel and
transverse field of a $Nb/Al_{ox}/Nb$ overlap-type junction with
$\lambda_J \sim 50 \mu m$ whose length is $500\mu m$, while the
width is equal to $4\mu m$. The base and top electrode widths are
$540$ and $506\mu m$, respectively. The junction geometry is
depicted in Fig.\ref{overexp}b. We observe that the two experimental
datasets almost overlap, when a factor scale of about $40$ is
applied on the abscissae, meaning that the sample is much more
sensitive to a transverse field rather than an in-plane one.

\subsection{Cross type junctions}

\begin{figure}[h]
\begin{center}

\epsfysize=6.0cm \epsfbox{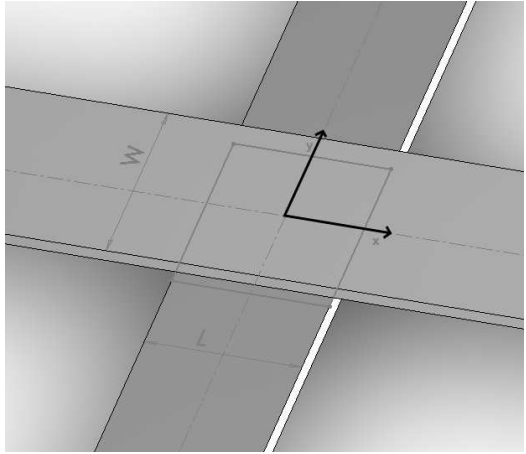}
\end{center}
\caption{Sketch of a square cross type junction. The center of the
junction coincides with the axis origin.} \label{crossgeom}
\end{figure}

\begin{figure}
\centering

\subfigure[]{\includegraphics[width=7cm]{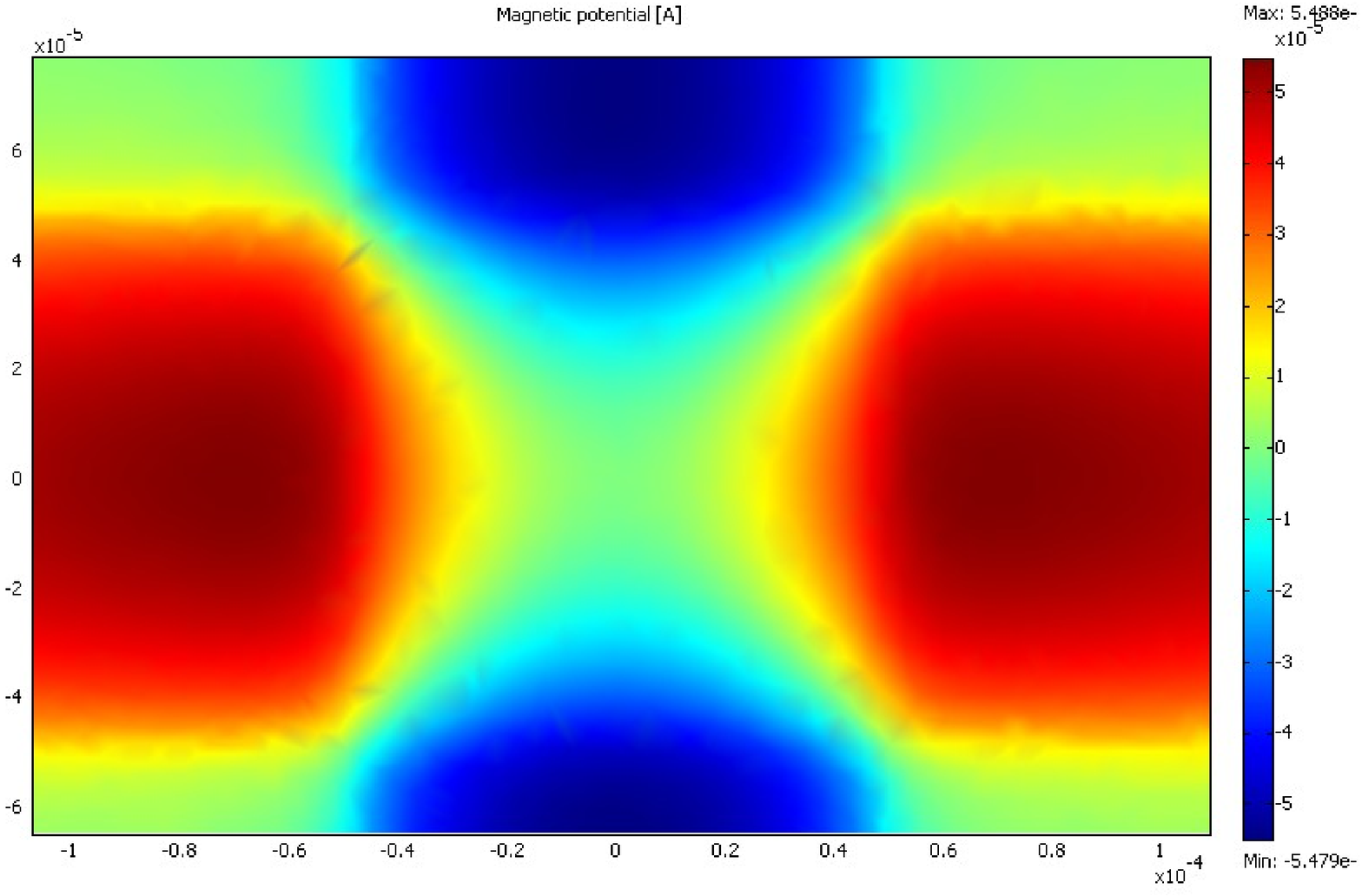}}%
\subfigure[]{\includegraphics[width=7cm]{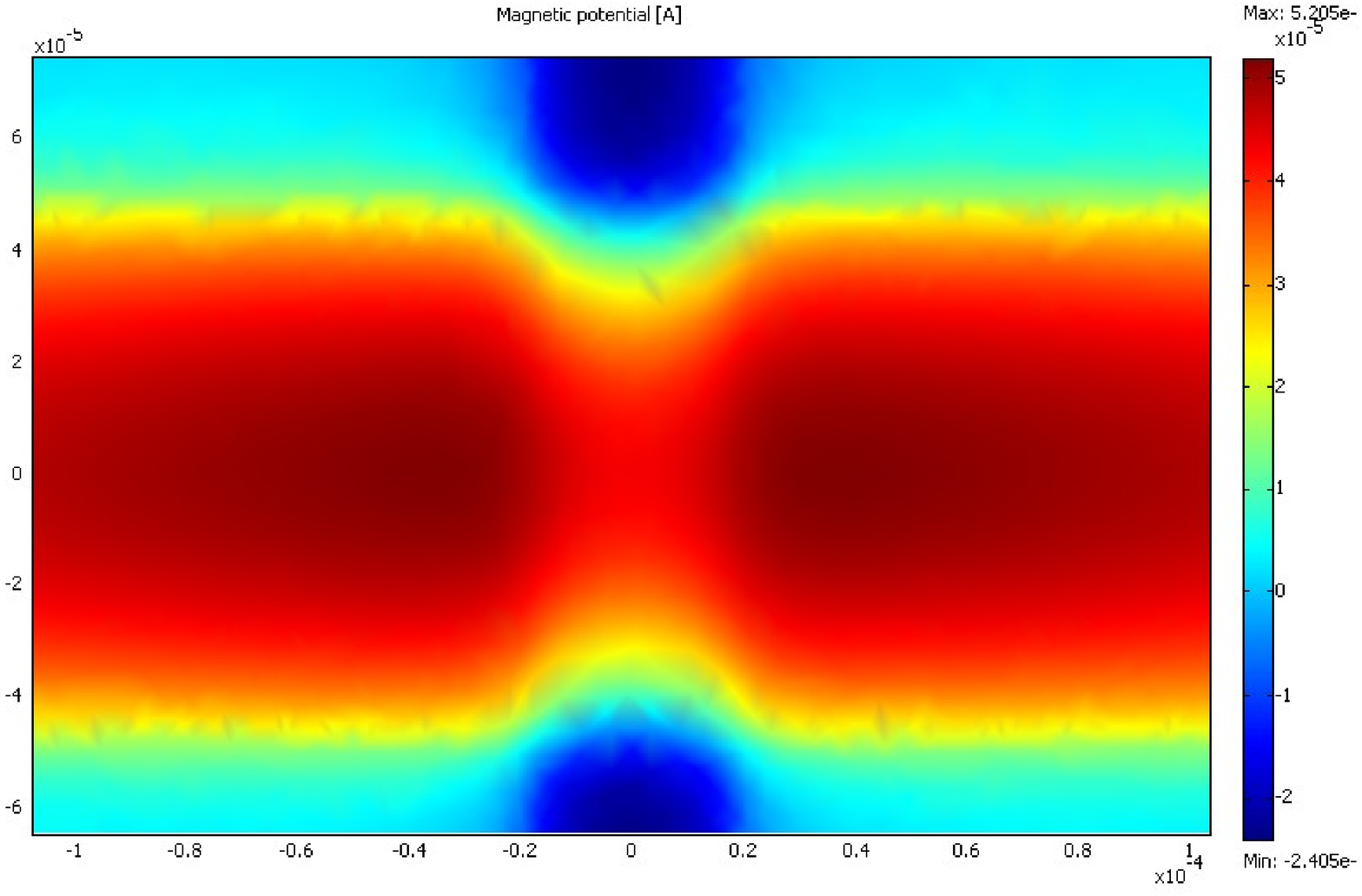}}
\subfigure[]{\includegraphics[width=7cm]{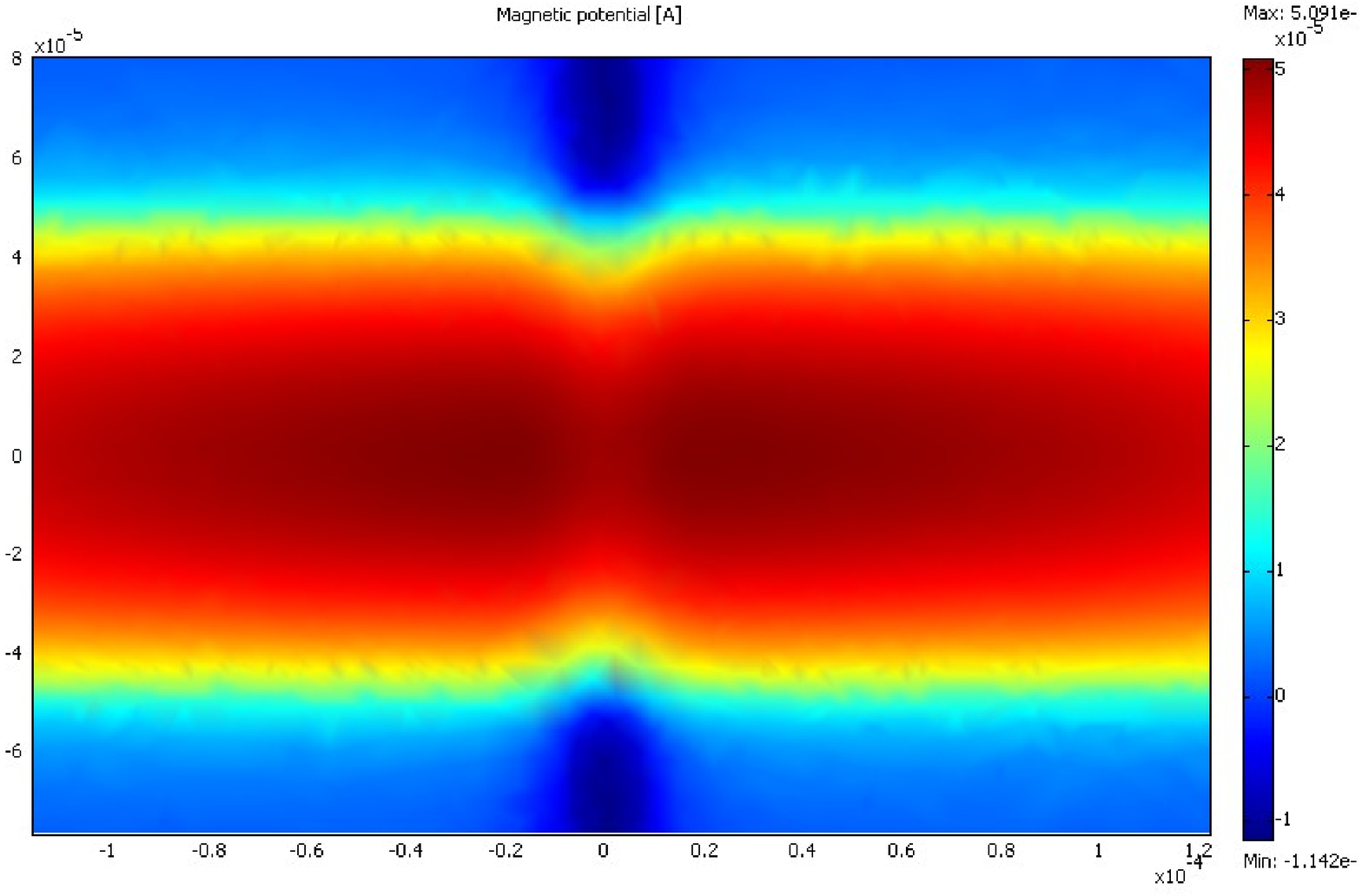}}

\caption{(Color online) Numerically obtained magnetic scalar
potential $V_m$ (in $A$) in the barrier plane for three cross type
junctions having the same width $2W=100\mu m$, but different
lengths: a) $2L=100\mu m$ ($\beta=1$), b) $2L=40\mu m$
($\beta=0.4$), and c) $2L=20\mu m$ ($\beta=0.2$). The external
applied field is $H_\bot=1A/m$.} \label{Vmcross}
\end{figure}

\noindent Cross geometry JTJs are formed by the superposition of two
perpendicular superconducting electrodes, as depicted in
Fig.\ref{crossgeom}. The static properties of such junctions in a
transverse magnetic field were analyzed by Miller \textit{et
al.}\cite{miller}, but only in the particular case of equal
 film widths $2L=2W$. They proposed, as an approximate solution of
Eq.\ref{laplace}, a phase profile $\phi(x,y)\propto xy$
(corresponding to a saddle shaped scalar magnetic potential
$V_m(x,y)\propto x^2-y^2$ and to a monotonically decreasing
$I_c(H_{\bot})$.) We want to generalized these results for junctions
with non unitary aspect ratios $\beta=L/W$. Figs.\ref{Vmcross}a-c
display the numerical solutions of Eq.\ref{Vm} for three cross
junctions having the same width $2W=100\mu m$, but different lengths
$2L=100$, $40$, and $20\mu m$. We observe that, for any value of
$\beta$, the scalar potential in the barrier is four-fold symmetric
meaning that the empirical expression $V_m(x,y)$ we are looking for
has to be an a even function of both $x$ and $y$. Further, $V_m$
always vanishes at the junction corners and, as the junction length
shrinks, the scalar potential distribution gets more and more
uniform over the barrier area. A careful analysis of the scalar
potential plots in Figs.\ref{Vmcross}a-c, led us to the following
expression:

\begin{equation}
\label{crossVm} V_m(\hat{x}, \hat{y}) = \hat{H} \sqrt{WL} \left(
\cos q \hat{y}\frac{\cosh q \beta \hat{x}}{\cosh\beta} - \cos q
\hat{x}\frac{\cosh q \hat{y}/\beta}{\cosh 1 / \beta} \right),
\end{equation}

\noindent in terms of normalized variables. Here again $q$ is a
fitting that can be comfortably set equal to unity. More
specifically, with $q=1$, the relative difference between the
simulation output and the heuristic expression of Eq.\ref{crossVm}
was numerically found to be everywhere less than $\pm 15 \%$
although it was minimized by $q\simeq1.2$. The proposed expression
is made up by two terms which can be seen as the contributions from
the two electrodes. When $L=W$, the two terms have the same weights
($\textrm{sech}1\simeq 0.65$), but, for, say, $\beta<1$, the weight
of the first term is larger than the one of the second term and
viceversa. Further, in the limit $\beta<<1$ the first weight
saturates to unity while the second vanishes.

\begin{figure}
\subfigure[]{\includegraphics[width=8.2cm]{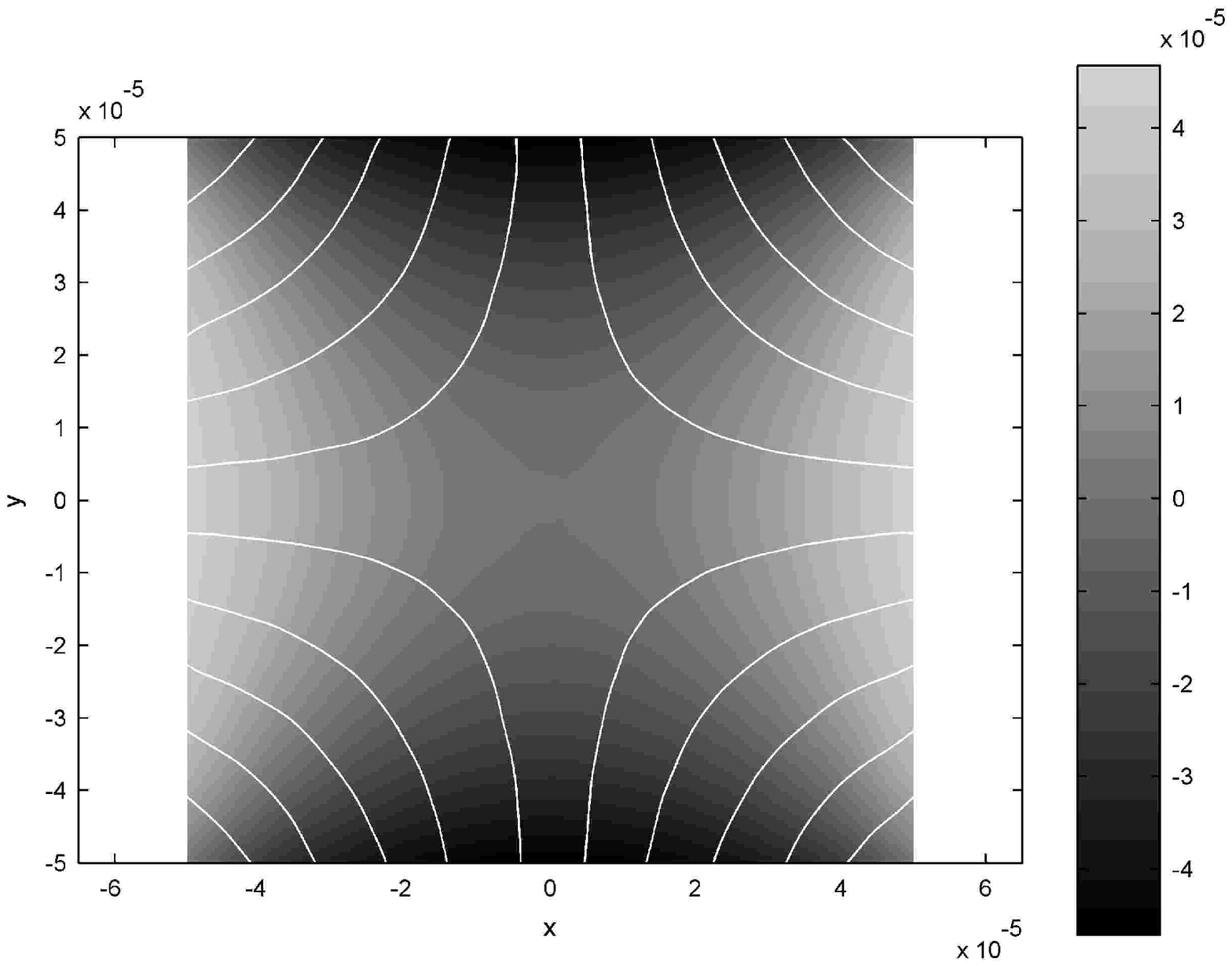}}
\subfigure[]{\includegraphics[width=7cm]{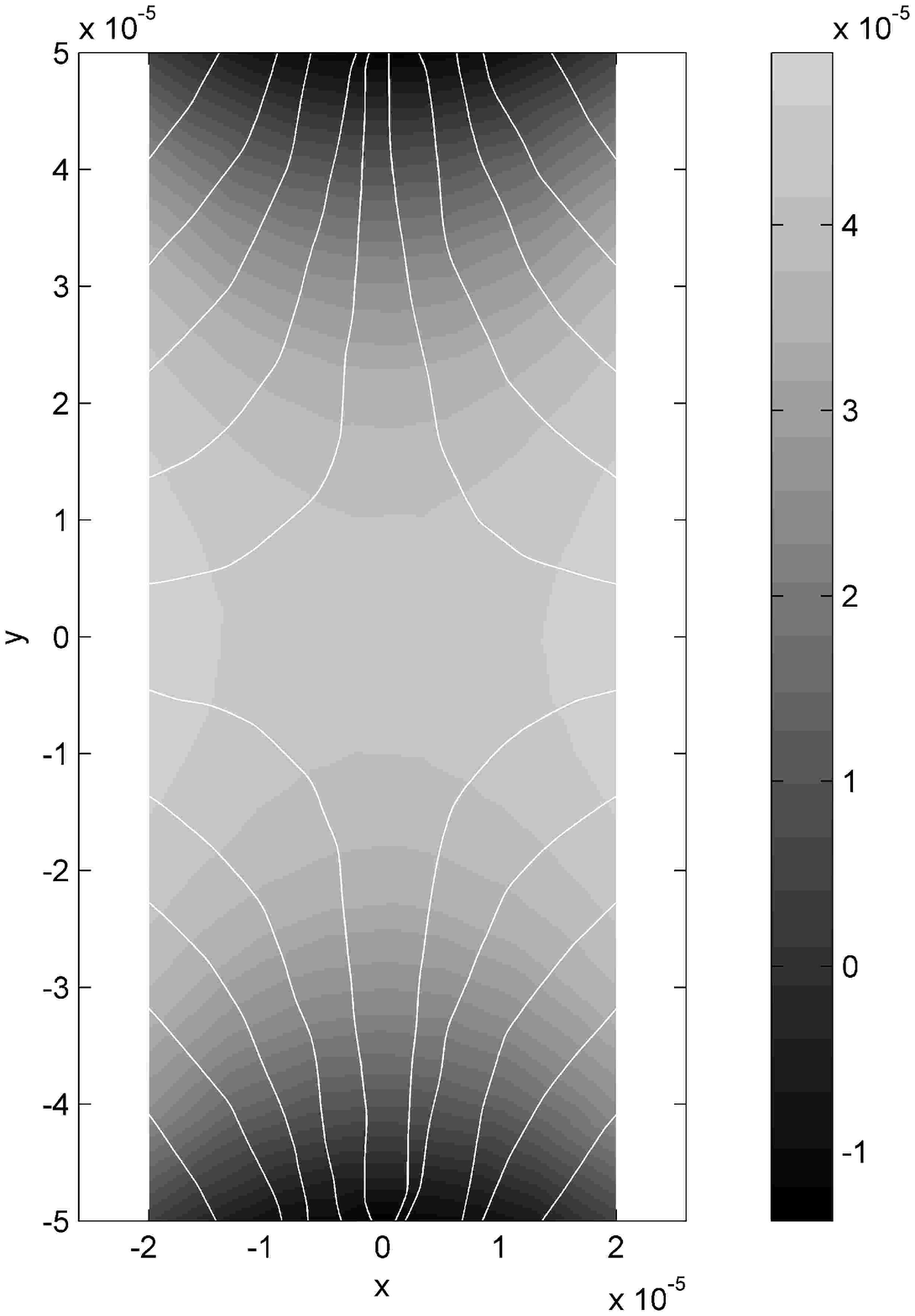}}%
\subfigure[]{\includegraphics[width=5cm]{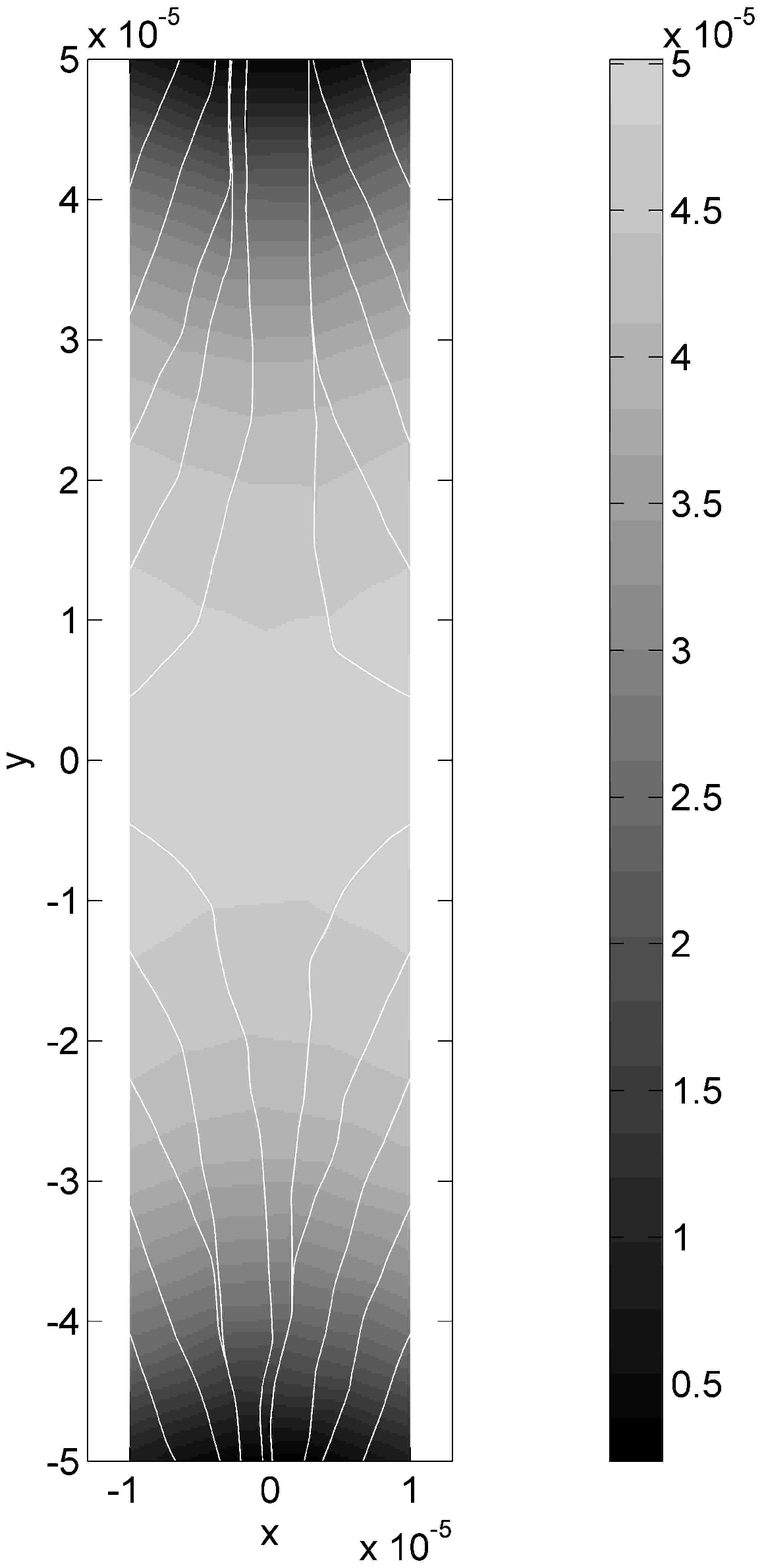}}
\caption{Numerically obtained magnetic field distribution in the
barrier area of three cross-type planar Josephson tunnel junctions
having the same width $2W=100\mu m$, but different lengths: a)
$2L=100\mu m$ ($\beta=1$), b) $2L=40\mu m$ ($\beta=0.4$) and c)
$2L=20\mu m$ ($\beta=0.2$).} \label{CrossStream}
\end{figure}

\noindent From Eq.\ref{crossVm} with unitary $q$, the Josephson
phase profile can be easily derived:

\begin{equation}
\label{phicross} \phi(\hat{x}, \hat{y}) = h \left(
\sin\hat{y}\frac{\sinh\beta \hat{x}}{\cosh\beta} +
\sin\hat{x}\frac{\sinh \hat{y}/\beta}{\cosh1/\beta}\right),
\end{equation}

\noindent where $h=2\pi d_e \sqrt{WL} \mu_0 \hat{H} /\Phi_0$ and
with Eq.\ref{laplace} being identically satisfied. We begin with the
observation that setting $\beta=1$ and retaining the first two terms
in the Taylor expansion of the trigonometric and hyperbolic
functions, Eqs.\ref{phicross} and \ref{crossVm} reduces to
$\phi(x,y)\propto xy - x^3 y^3/36 $ and $V_m(x,y)\propto x^2 - y^2$,
as it should be. Further, upon the inversion of $\beta$,
$\phi(\hat{x}, \hat{y})=\phi(\hat{y}, \hat{x})$, meaning that the
solutions for two junctions having reciprocal aspect ratios differ
by a rotation of $\pm 90^o$. Figs.\ref{CrossStream}a-c show the
magnetic field distributions in the barrier area corresponding to
the scalar potentials shown in Figs.\ref{Vmcross}a-c.

\begin{figure}[htp]
\centering \subfigure{\includegraphics[width=8cm]{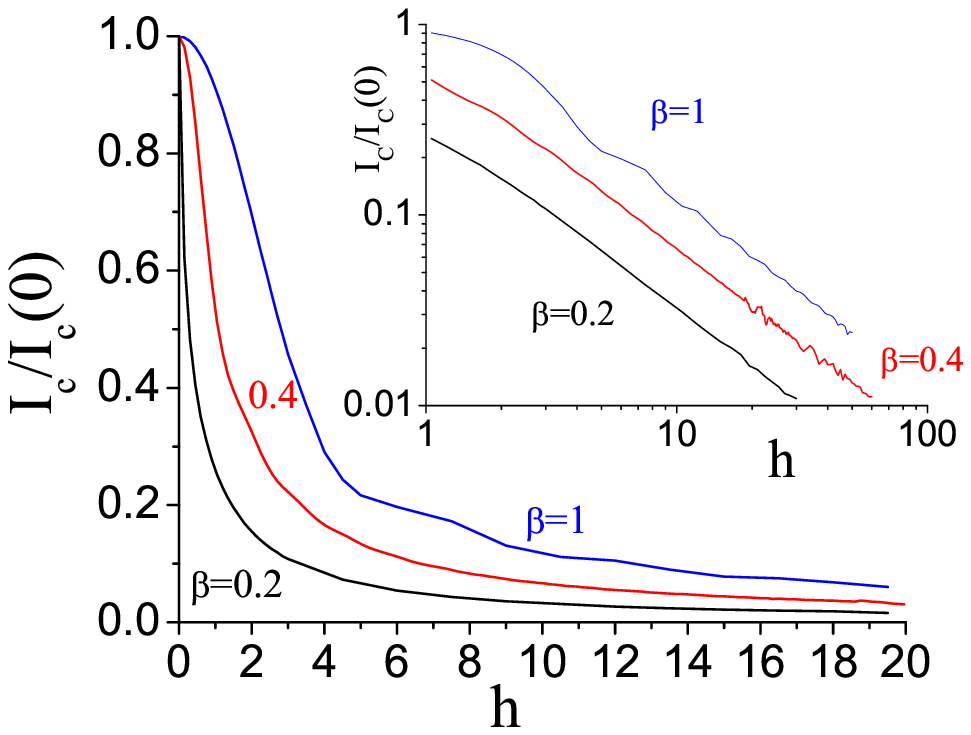}}

\caption{(Color online) Computed transverse magnetic patterns
$I_c(h)$ for a cross junction with different $L/W$ ratios $\beta=1$,
$\beta=0.4$ and $\beta=0.2$. In the inset the log-log plot shows
that for large fields $I_c(h)\propto 1/h$.} \label{pattcross}
\end{figure}

\noindent Again, with $\langle \sin \phi \rangle=0$, the magnetic
diffraction pattern for a cross junction in a transverse magnetic
field are found on inserting the expression above in
Eq.\ref{overpatt}. Fig.\ref{pattcross} shows $I_c(h)$  for the three
values of the barrier aspect ratios considered in
Figs.\ref{Vmcross}a-c, i.e., $\beta=1$, $0.4$, and $0.2$. For the
considerations above, the red and black  curves in
Fig.\ref{pattcross} also represent the $I_c(h)$ for $\beta=2.5$ and
$5$, respectively. We come to the interesting result that for cross
junctions in a transverse field the critical current decreases
monotonically with the field amplitude $H_{\bot}$ and, for large
fields ($h>>1$), $I_c(h)\propto 1/ H_{\bot}$ (see the log-log plot
in the inset of Fig.\ref{pattcross}). The experimental transverse
pattern presented in Ref.\cite{miller} bears strong resemblance to
the $I_c(h)$ obtained for $\beta=1$.

\section{Annular junctions}

\noindent In this section we will examine the behavior of small
annular JTJs in the presence of a transverse magnetic field.
Denoting the inner and outer ring radii, respectively, as $r_i$ and
$r_o$, we assume that the annular junction is unidimensional, i.e.,
the ring mean radius $\overline{r}= (r_i+r_o)/2$ is much larger than
the ring width $\triangle r = (r_o-r_i)/2$.

\begin{figure}[htp]
\centering

\subfigure[]{\includegraphics[width=8cm]{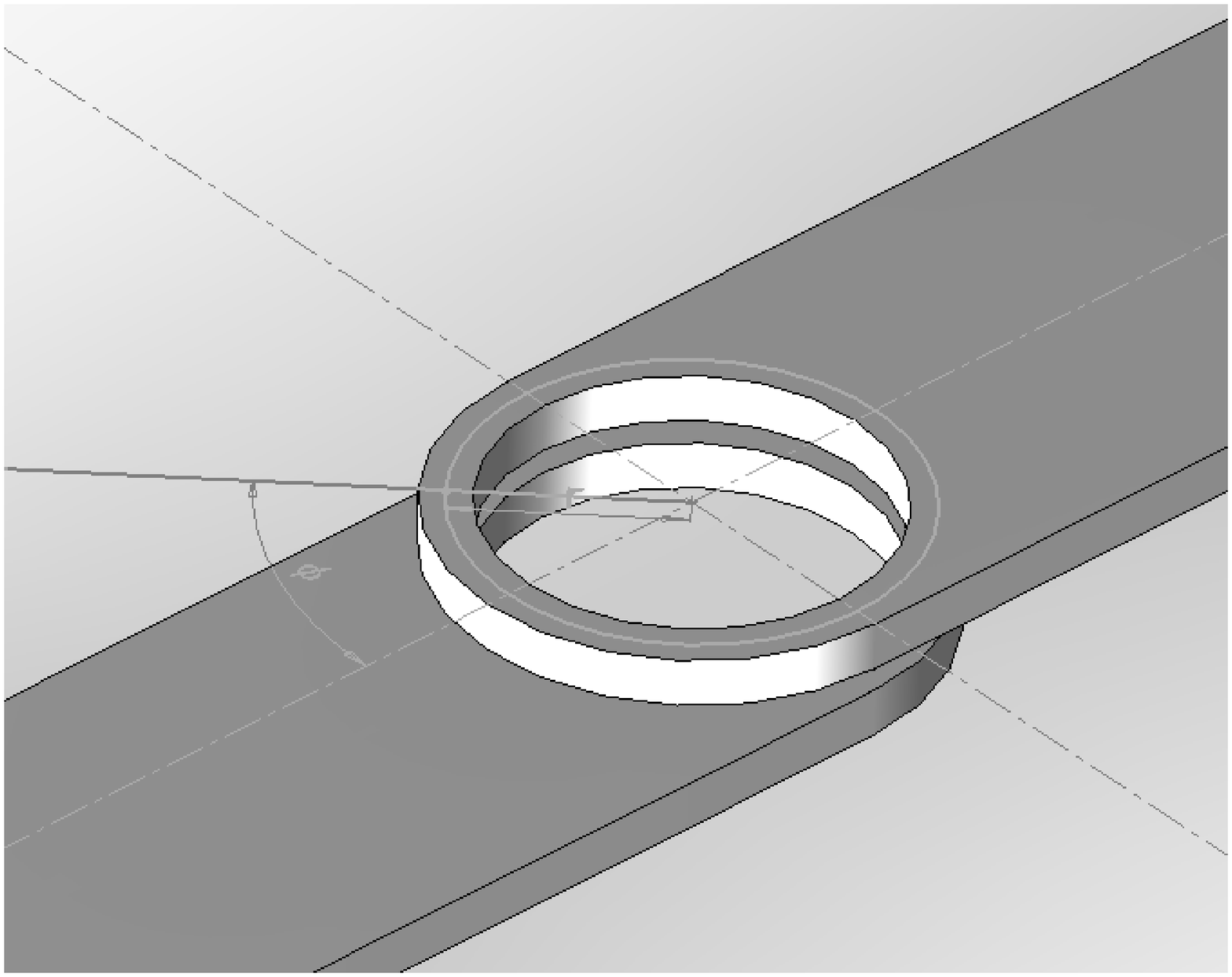}}
\subfigure[]{\includegraphics[width=8cm]{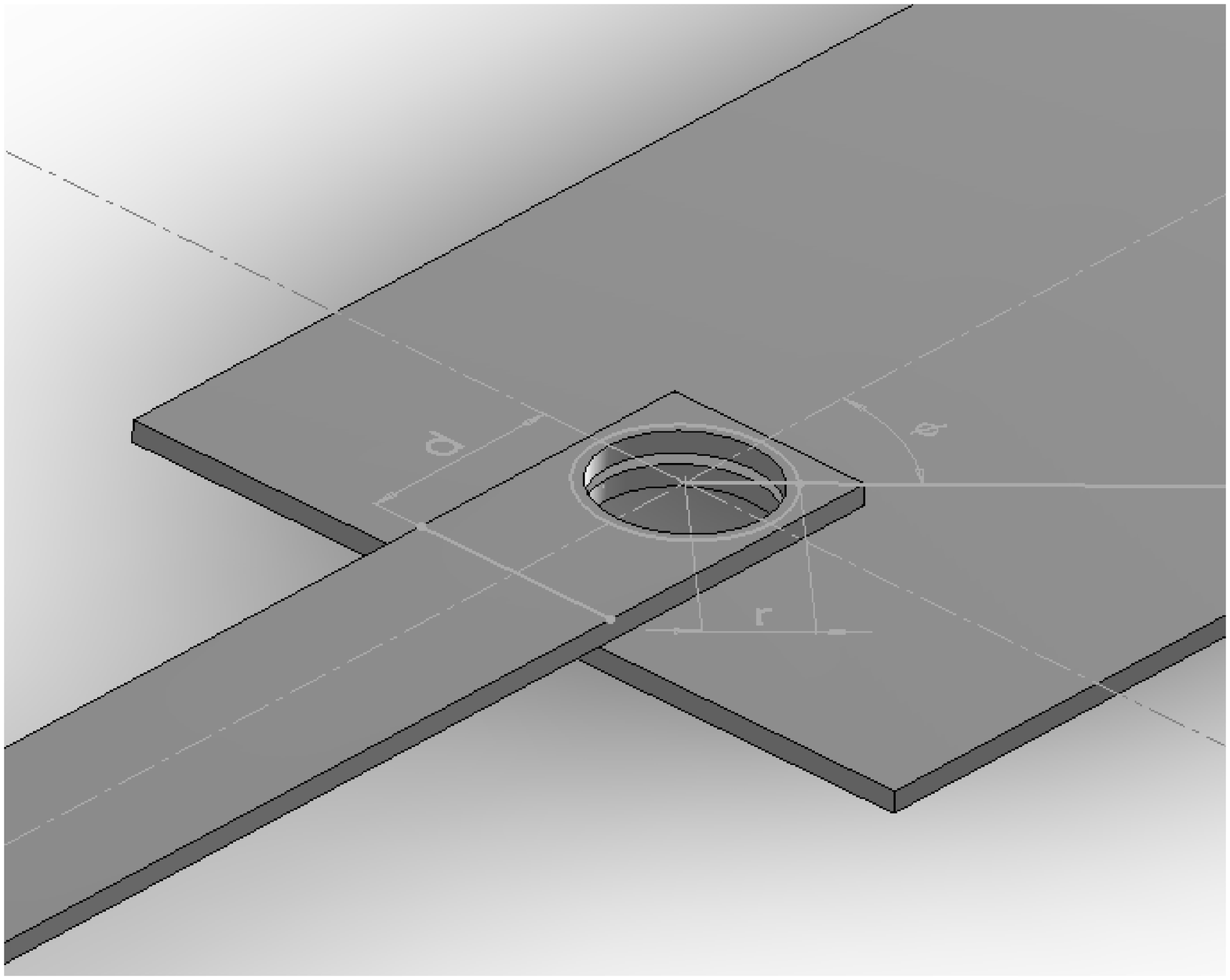}}

\caption{(Color online) Sketches of the two types of annular JTJs
considered in this paper. a) \textit{Lyngby} type geometry made by
two films whose widths match the ring diameter. b) Asymmetric
annular junctions made by two films with unequal widths.}
\label{anngeoms}
\end{figure}

\noindent Using polar coordinates, the Josephson equation
Eq.\ref{gra} can be split into:

\begin{equation}
\label{Josann} {{\partial \phi} \over {\partial r}} =\kappa H_\theta
\quad  ,\,\,\,\, \quad
  {{\partial \phi} \over {r \partial \theta}} =-\kappa H_r ,
\end{equation}

\noindent where $H_r$ and $H_\theta$ are the radial and tangential
components of the magnetic field in the ring plane, respectively and
$\kappa$ depends on the electrodes geometrical
configuration\cite{prb96}. With the annulus unidimensional, we can
neglect the radial dependence of the Josephson phase
$\phi(r,\theta)=\phi(\overline{r},\theta)$ and, henceforth:

\begin{equation}
\label{phiann} \phi=\phi(\theta)=-\kappa \overline{r} \int d\theta
H_r(\overline{r},\theta) + const.
\end{equation}

\noindent In the well known case of a spatially homogeneous in-plane
field $\bf{H_{\|}}$ applied in the direction of $\theta=0$, then
$H_r=H_{\|}\ \cos\theta$ (and $H_r=H_{\|}\ \sin\theta$), so that the
last integral yields\cite{br96}:

\begin{equation} \label{small}
\phi(\theta) = h \sin \theta + \phi_0 ,
\end{equation}

\noindent where $h \propto H_{\|}$ and $\phi_0$ is an integration
constant. Assuming that the \Jos current density $J_c$ is constant
over the ring circumference, the \Jos current through the barrier is
obtained by:

\begin{equation} \nonumber
I_c(h)=\frac{I_c(0)}{2\pi}\int_{-\pi}^{\pi} d\theta \sin
\phi(\theta).
\end{equation}

\noindent in which $I(0)=J_c 2\pi {\bar r}\Delta r$ is the maximum
\jun critical current which occurs in zero field. As far as
$\phi(\theta)$ is an odd function (when $\phi_0=0$), the calculation
of the maximum critical current reduces to the following
integration:

\begin{equation}
\label{annpatt} I_c(h)=\frac{I_c(0)}{\pi}\int_{0}^{\pi} d\theta \cos
\phi(\theta).
\end{equation}

\noindent Inserting $\phi$ given by Eq.\ref{small}, we obtain for
the maximum critical current\cite{prb96},

\begin{equation} \label{Ic0}
I_c(h)= I_c(0) \bigl| J_0(h)\bigr|,
\end{equation}

\noindent  in which $J_0$ is the zero order Bessel function (of
first kind). The periodic conditions for the phase difference $\phi$
and its angular derivative around an annular \jun are:

\begin{equation} \label{peri1}
\phi(\theta + 2 \pi)=\phi(\theta)+ 2\pi n,
\end{equation}
\begin{equation} \label{peri2}
{{d \phi(\theta+2\pi) } \over { d \theta}} = {{d \phi(\theta) }
\over { d \theta}},
\end{equation}

\noindent where $n$ is an integer corresponding to the net number of
fluxons (i.e., number of fluxons minus number of antifluxons)
trapped in the \jun at the time of the normal-to-superconducting
transition. Eqs.(\ref{peri1}) and (\ref{peri2}) state that
observable quantities such as the \Jos current (through $\sin\phi$)
and the radial magnetic field (through $d \phi  / d \theta$) must be
single valued upon a round trip; they were derived in
Ref.\cite{prb96} starting from the fluxoid quantization.

\noindent Eq.\ref{small} and Eq.\ref{Ic0} hold under the assumption
that there are no fluxons trapped in the barrier; however, they can
be easily generalized to the case of $n\neq 0$ trapped fluxons. In
such case, Eq.\ref{small} changes to:

\begin{equation} \label{smal}
\phi(\theta) = h \sin \theta + n\theta + \phi_0,
\end{equation}

\noindent in which the linear term in Eq.\ref{smal} takes into
account the phase twist due to the presence of the trapped fluxons,
being that the ring circumference is smaller or comparable to the
fluxon rest length. Carrying out the integration in Eq.\ref{annpatt}
with $\phi$ given by Eq.\ref{smal} and maximizing with respect to
$\phi_0$, we get:

\begin{equation} \label{Icg}
I_c^n(h)= I_c(0) \bigl| J_n(h)\bigr|,
\end{equation}

\noindent in which $J_n$ is the n-th order Bessel function.
Eqs.\ref{Ic0} and \ref{Icg} have been experimentally verified in a
number of papers.

\noindent A \textit{Lyngby} type annular JTJ, firstly reported in
1985 by Davidson \textit{et al.}\cite{davidson}, is obtained by two
films having the same width, as schematically depicted in
Fig.\ref{anngeoms}a. Further, Fig.\ref{anngeoms}b shows a different
kind of annular JTJ for which the film widths are quite different:
we will call it  \textit{asymmetric annular junction}. At the end of
this section we will present experimental data for such asymmetric
geometrical configuration. We have carried out magnetostatic
simulation for the two annular geometries depicted in
Figs.\ref{anngeoms} when the applied field is transverse. Only the
case of no trapped fluxons was considered, corresponding to zero net
magnetic flux through the superconducting holes. Contrary to the
case of the rectangular bidimensional JTJs considered previously,
now we do not have to know the magnetic field distribution in the
junction plane, but, by virtue of Eq.\ref{Josann}b, we can limit our
interest to just the angular dependence of the radial magnetic field
$H_r(\overline{r},\theta)$. In our simulations we set $r_i=40\mu m$
and $r_o=50 \mu m$, so that $\overline{r}= 45 \mu m$. For the
asymmetric configuration the film widths were chosen to be
$2W=100\mu m$ and $2W'=200\mu m$.

\begin{figure}[htp]
\centering

\subfigure[]{\includegraphics[width=8cm]{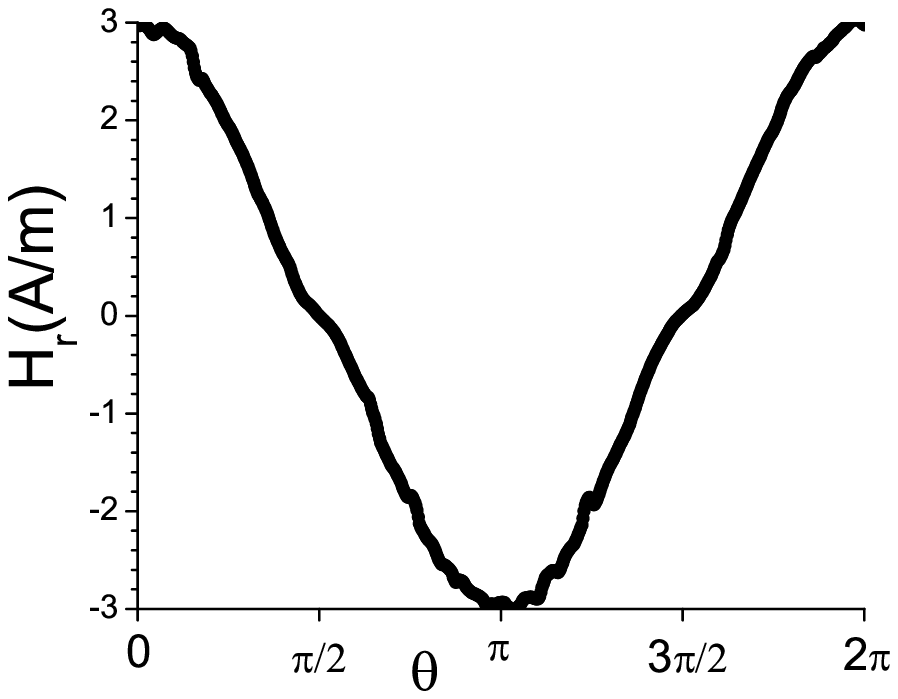}}
\subfigure[]{\includegraphics[width=8cm]{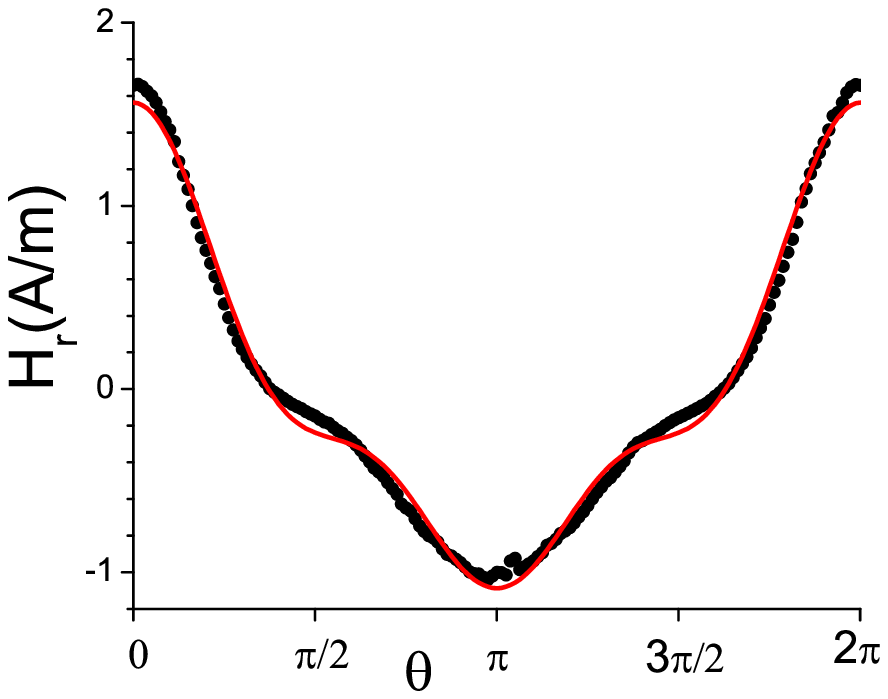}}

\caption{(Color online) Angular dependencies of the radial magnetic
field $H_r(\theta)$ for a small annular junction in a transverse
field having: a) \textit{Lyngby} type geometry sketched in
Fig.\ref{anngeoms}a and b) asymmetric geometry sketched in
Fig.\ref{anngeoms}b. In the numerical simulations the amplitude of
the transverse field was set to $1A/m$.} \label{phitheta}
\end{figure}

\noindent Postprocessing the simulation outputs we found out that,
in the case of Lyngby geometry, $H_r$ follows very closely a
sinusoidal dependence on $\theta$, as shown in Figs.\ref{phitheta}a:
more specifically, by choosing the angle origin in such a way that
$\theta=0$ corresponds to the positive x-axis direction, we have
$H_r \propto \cos \theta$, exactly as if the magnetic field were
applied in the ring plane. By integration we get Eq.\ref{smal} again
with $h$ depending on the geometrical film configuration and being
proportional to the transverse field amplitude $H_\bot$. We come to
the remarkable conclusion that the diffraction pattern of an
electrically small annular junction with no trapped flux in a
transverse field follows the zero order Bessel function behavior, as
if the field were applied in the barrier plane.

\noindent The situation is quite different when we consider
asymmetric annular junctions. In fact, as shown in
Fig.\ref{phitheta}b, it is quite evident that now the slope of the
radial field changes abruptly for $\theta \simeq \pi/2$ and $\theta
\simeq 3\pi /2$ resulting in a periodic asymmetric ratchet-like
potential $d H_r/d\theta$. We have numerically checked that to a
high accuracy $\int_0^{2\pi} d \theta H_r(\theta)=0$, as it should
be when no fluxons are trapped in the junction. In order to
correctly reproduce $H_r(\theta)$, we have to consider higher
$\theta$ harmonics. It was found that a truncated Fourier expansion
cast in the form:

\begin{equation} \label{Hradial}
H_r(\theta) \propto \cos \theta + 2 \gamma  \cos 2\theta + 3 \delta
\cos 3\theta
\end{equation}

\noindent can satisfactorily fit our numerical findings. The two
fitting parameters $\gamma$ and $\delta$ can be ascribed to two
degrees of freedom in the layout geometry: the ratio of the top and
bottom film widths and the distance from the junction to the edge of
the bottom film. Eq.\ref{Hradial} with $\gamma=0.11$ and
$\delta=0.085$ is shown as a solid red line in Fig.\ref{phitheta}b.

\begin{figure}[htp]
\centering

\subfigure[]{\includegraphics[width=8cm]{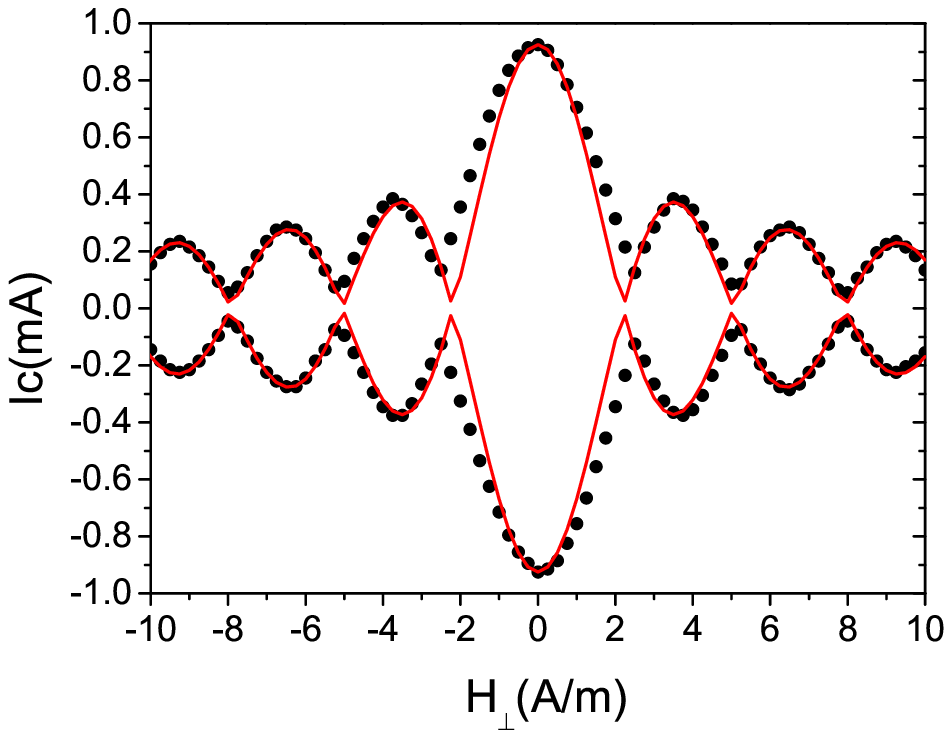}}
\subfigure[]{\includegraphics[width=8cm]{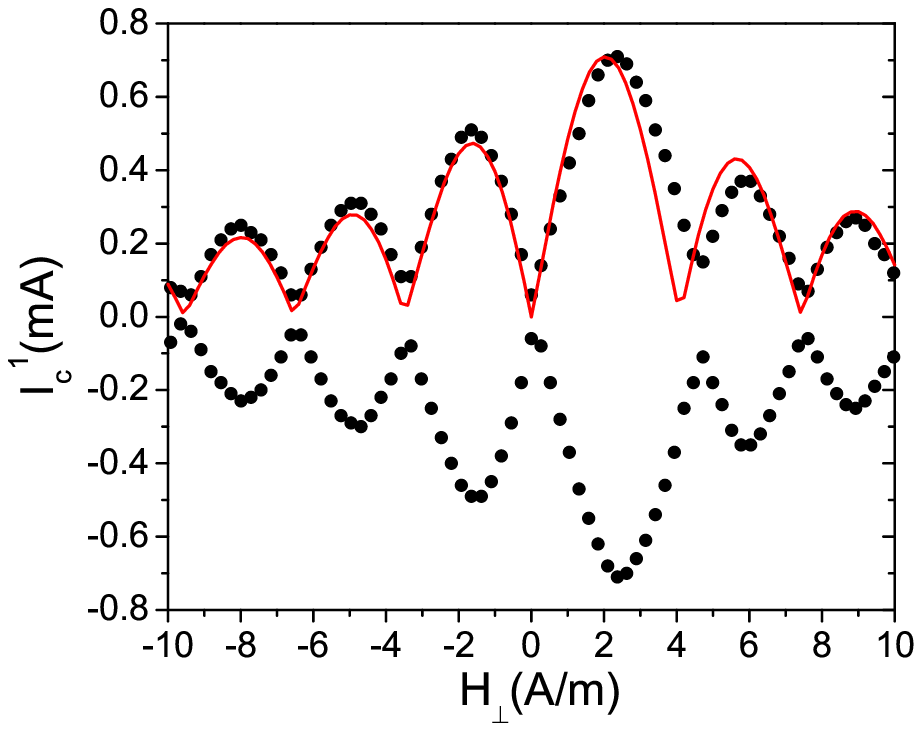}}

\caption{(Color online) Transverse diffraction patterns for an
asymmetric annular junction. The dots are the experimental data,
while the solid red lines are the theoretical expectations obtained
inserting $\gamma=0.18$ and $\delta=0.08$ in Eq.\ref{annular}. a) no
trapped fluxon; b) one trapped fluxon.} \label{xy}
\end{figure}

\noindent By integrating Eq.\ref{Josann}b with $H_r$ given by
Eq.\ref{Hradial}, we get an approximate expression for the angular
phase dependence:

\begin{equation} \label{annular}
\phi(\theta) \approx  h ( \sin \theta + \gamma  \sin 2\theta +
\delta \sin 3\theta)
\end{equation}

\noindent in which still $\phi(-\theta)=-\phi(\theta)$, but the
symmetries $\phi(\pi/2-\theta)=\phi(\pi/2+\theta)$ and
$\phi(3\pi/2-\theta)=\phi(3\pi/2+\theta)$ are now lost. Since
$\phi(\theta)$ is an odd function, Eq.\ref{annpatt} allows us to
calculate the magnetic diffraction patterns corresponding to the
above non-sinusoidal phase profile, even in the case when a term
$n\theta$ is added to account for the presence of $n$ trapped
fluxons. It turned out that, while for $n=0$ the effects of the
$\gamma$ and $\delta$ terms tend to cancel each other, resulting in
a zero-order Bessel function behavior as in Eq.\ref{Ic0}, for
$n\neq0$ we found a marked departure from the $n$-th order Bessel
function dependence of Eq.\ref{Icg}. These results are supported by
experimental results for an asymmetric annular junction
($\overline{r}= 80 \mu m$ and $\Delta r= 4 \mu m$) made by unequal
width films: the base electrode width is $540 \mu m$ and the top
electrode width is $170 \mu m$. For such layout, the numerical
analysis of the angular radial field dependence yielded the best fit
values $\gamma=0.19$ and $\delta=0.078$. In the Figs.\ref{xy}a and b
we show, respectively, the experimental diffraction patterns (dots)
for such junction without trapped fluxons and with $n=1$ trapped
fluxon. The experimental data can be fitted very nicely by the
theoretical expectations (solid red lines) obtained inserting the
above $\gamma$ and $\delta$ values in Eq.\ref{annular}.

\noindent We observe that when no fluxons are trapped in the
asymmetric annular junctions the transverse pattern is definitely
symmetric with respect to the inversion of field direction and is
barely distinguishable from the pure Bessel one; further, we stress
that the same sample measured with an in-plane field applied in the
$\theta=0$ direction showed again a Bessel like pattern, but the
response to the applied field was about 25 times weaker.

\noindent On the contrary, with $n=1$ the transverse magnetic
diffraction pattern loses its symmetry with respect to the field
amplitude, i.e. $ I_c^1(-h) \neq I_c^1(h)$. Furthermore, both in the
experiments and in the calculations, it turns out that $ I_c^1(-h) =
I_c^{-1}(h)$; in other words, if we invert both the field and fluxon
polarities we obtain the same magnetic diffraction pattern. This
result was obtained and exploited in the context of a detailed
investigation of the symmetry breaking during fast
normal-to-superconducting phase transitions of annular JTJs recently
published\cite{KZ}. Among other things, it has been experimentally
and theoretically demonstrated that when a small transverse field is
applied to the ring during the thermal quench the probability to
trap a Josephson fluxon can be very close to unity, the fluxon
polarity depending on the field polarity. The ability to easily
discriminate between a fluxon and an antifluxon can be conveniently
exploited in the recently proposed Josephson-vortex qubits
experiments with ring and heart-shaped JTJs\cite{clarke}. The
asymmetry of the magnetic diffraction pattern can be very simply
ascribed to the ratchet-like potential whose effect on the fluxon
dynamic properties has been fully investigated
recently\cite{goldobin}.

\noindent We conclude this section by remarking that the angular
dependence of both the radial and tangential magnetic field
components in the barrier of a annular JTJ do not change if the
circular hole is removed from one of the electrodes. This is
supported by both numerical simulations and experimental
data\cite{prb98}. Indeed, when the ring shaped barrier is formed
between a holed film and a singly connected one, the Josephson
fluxon polarity is univocally related to the polarity of the
quantized flux threading the hole.

\section{Concluding remarks}

\noindent The transverse magnetic patterns of electrically small
Josephson tunnel junctions have been derived numerically by solving
the magnetostatic problem for different geometrical configurations
of the junction electrodes and of the barrier. More specifically,
from the numerical analysis of the magnetic scalar potential
produced in the barrier plane by the demagnetizing currents
circulating on the electrode surfaces we derived approximate and
simple expressions for the Josephson phase distribution in the
barrier area, which, in turn, permitted to calculate the junction
critical current. Such calculations show, among other things, that
for rectangular barriers the modulation of the maximum critical
current never follows the Fraunhofer behavior typical of a field
applied in the barrier plane; further, $I_c(H_{\bot})$ strongly
depends on the barrier aspect ratio $L/W$. On the contrary, the
critical current modulation in a transverse field of annular JTJs
without trapped fluxons is fairly close to the one corresponding to
a parallel field, although it can be much faster when the field is
perpendicular. When the film configuration of the annular junction
is asymmetric, then the static properties depend on the polarity of
the transverse field and of the trapped fluxons. It's worthy to
mention that our calculations were carried out assuming that the
junctions were not biased. However, in order to measure the magnetic
diffraction patterns one needs to supply a transport current by an
external source. As far as the JTJ is electrically small, as in
cases considered in this paper, the effect of a non uniform current
distribution through the barrier is negligible\cite{barone}.
Nevertheless, to exclude flux from the electrodes interiors, a
self-field that wraps around the films is generated\cite{Rose-Innes}
whose effect on the Josephson phase distribution is largest when the
current is largest. This situation typically occurs when the applied
field is small (or absent) regardless of its orientation with
respect to the barrier plane. As the external field amplitude grows,
the relative effect of bias induced screening currents decreases,
and disappears when the field amplitude is such that the critical
current is zero.

\noindent It is important to stress that the static properties of a
small JTJ in a transverse field is strictly related to the film
layout. In the case of junctions formed in a windows between two
films which completely overlap each other near the junction itself,
the circulating currents on the film interior surfaces are symmetric
with respect to the barrier plane and result in a zero magnetic
field; consequently such JTJs will remain totally insensitive to a
transverse field: this holds for overlap type and annular geometry
JTJs shown, respectively, in Fig.\ref{overgeom} and
Fig.\ref{anngeoms}a when one of the electrodes is rotated by
$180^o$. As mentioned in the Introduction, we also expect a very
small sensitivity to a transverse field when the barrier window is
located well inside the superconducting electrodes. The possibility
to design multijunction chips whose each junction has its own
magnetic diffraction pattern makes the physics and the application
of transverse field very attractive and promising. Unfortunately, so
far, very few experimental works have dealt with transverse field
because of lack of theoretical understanding. We believe that this
paper will stimulate other groups to fill the gap.


\end{document}